\documentclass[12pt,preprint,showpacs]{revtex4}
\usepackage{amsmath,amsthm,amssymb,amsfonts}
\usepackage{graphicx}

\newtheorem{lemma}{Lemma}
\newtheorem{proposition}{Proposition}

\newtheorem{theorem}{Theorem}

\newcommand{\C}{\mathbb{C}}
\renewcommand{\epsilon}{\varepsilon}
\newcommand{\E}{\mathcal{E}}
\newcommand{\Eh}{\widehat{\mathcal{E}}^{\rm M}}
\newcommand{\eh}{\widehat{{E}}^{\rm M}}

\renewcommand{\phi}{\varphi}
\newcommand{\R}{\mathbb{R}}
\newcommand{\ab}{\mathbf{a}}
\newcommand{\x}{\mathbf{r}}
\newcommand{\y}{{\mathbf{r}'}}
\newcommand{\s}{\mathbf{s}}
\newcommand{\tb}{\mathbf{t}}
\newcommand{\X}{\mathbf{x}}
\newcommand{\z}{\mathbf{z}}
\newcommand{\p}{\mathbf{p}}
\newcommand{\I}{\mathbb{I}}
\newcommand{\g}{\gamma}
\newcommand{\be}{\begin{equation}}
\newcommand{\RR}{\mathbf{R}}
\newcommand{\half}{\mbox{$\frac12$}}

\newcommand\bra\langle
\newcommand\ket\rangle
\renewcommand{\ker}{{\mathcal N} \! }

\def\tr{\mathop{\mathrm{tr}}\nolimits} 

\def\Xint#1{\mathchoice
{\XXint\displaystyle\textstyle{#1}}%
{\XXint\textstyle\scriptstyle{\raise 2pt\hbox{\tiny{$#1$}}}}%
{\XXint\scriptstyle\scriptscriptstyle{#1}}%
{\XXint\scriptscriptstyle\scriptscriptstyle{#1}}%
\!\int}
\def\XXint#1#2#3{{\setbox0=\hbox{$#1{#2#3}{\int}$}
\vcenter{\hbox{$#2#3$}}\kern-.5\wd0}}

\begin{document}

\title{M\"uller's Exchange-Correlation Energy in Density-Matrix-Functional
Theory}

\author{Rupert L. Frank}
\email{rupert@math.kth.se}
\affiliation{Department of Mathematics, Royal Institute
           of Technology, 100 44 Stockholm, Sweden}

\author{Elliott H. Lieb}
\email{lieb@princeton.edu}
\affiliation{Departments of Mathematics and Physics,
Princeton University,
           P.~O.~Box 708, Princeton, NJ 08544, USA}

\author{Robert Seiringer}
\email{rseiring@princeton.edu}
\affiliation{Department of Physics, Princeton University,
        P.~O.~Box 708,
        Princeton, NJ 08544, USA}

\author{Heinz Siedentop}
\email{h.s@lmu.de}
\affiliation{Mathematisches Institut,
        Ludwig-Maximilians-Universit\"at M\"unchen, 
Theresienstra\ss e 39, 80333 M\"unchen, Germany}

\date{September 28, 2009}


\begin{abstract}     
    The increasing interest in the M\"uller density-matrix-functional
    theory has led us to a systematic mathematical investigation of its
    properties. This functional is similar to the Hartree-Fock
    functional, but with a modified exchange term in which the square of
    the density matrix $\gamma(\X,\X')$ is replaced by the square of
    $\gamma^{1/2}(\X,\X')$.  After an extensive introductory discussion
    of density-matrix-functional theory we show, among other things,
    that this functional is convex (unlike the HF functional) and that
    energy minimizing $\gamma$'s have unique densities $\rho(\x)$, which
    is a physically desirable property often absent in HF theory.  We
    show that minimizers exist if $N \leq Z$, and derive various
    properties of the minimal energy and the corresponding
    minimizers. We also give a precise statement about the equation for
    the orbitals of $\gamma$, which is more complex than for HF theory.
    We state some open  mathematical questions about the theory
    together with conjectured solutions.
\end{abstract}

\maketitle

\tableofcontents

\section{Introduction}

The basic goal of density-functional theory is to express the energy
of a quantum-mechanical state in terms only of its one-particle
density $\rho(\x)$ and then to minimize the resulting functional (the
`density functional') with respect to $\rho(\x)$ (under the subsidiary
condition that $\int_{\R^3} \rho(\x) d\x =N=$ number of electrons) in
order to calculate the ground-state energy of the system, which could
be an atom or a molecule or a solid. Although the first -- and by far
most used and important density functional in theory, computation, and
mathematical investigation of multi-electron systems -- is the
Thomas-Fermi functional (Lenz \cite{Lenz1932}), strong interest in the
subject was triggered by Hohenberg and Kohn \cite{HohenbergKohn1964}.
We refer the reader interested in the recent developments to the books
by Eschrig \cite{Eschrig2003} and Gross and Dreizler
\cite{GrossDreizler1995} and the review \cite{Lieb1983D}.

While this program is possible in principal, experience has shown that
it is far from easy to guess the appropriate functional -- especially
if one wants the functional to be universal and not simply `tuned' to
the particular kind of atom or molecule under investigation. There are
also pitfalls connected with the admissible class of functions to use
in the variational principle \cite{Levy1982,Lieb1983D}.

Whereas the external potential energy can easily be expressed in terms
of the one-particle density, it is not known how to express the
kinetic energy and the interaction energy in terms of $\rho(\x)$.
Going from density- to density-matrix-functional theory eliminates the
first problem altogether, since all expectations of one-particle
operators can be expressed in term of the one-particle density matrix.
The density matrix analogue of the Hohenberg-Kohn density-functional
program was established by Gilbert \cite{Gilbert1975}. See also
\cite{Levy1979}.

The most difficult component of the density-functional to estimate is 
the exchange-correlation energy (which we shall henceforth simply call
exchange energy), and it is that energy that will concern us here.
Owing to this and other difficulties, it has been the tendency
recently to replace the energy as a functional of $\rho(\x)$ by a
functional of the one-body density {\it matrix}, $\gamma(\X, \X')$. In
this way it is hoped to have more flexibility and achieve, hopefully,
more accurate answers.

Fermions have spin and it 
is convenient to write a particle's coordinates as
$\X=(\x,\sigma)$ for a pair consisting of a vector $\x$ in space and
an integer $\sigma$ taking values from $1$ to $q$. Here $q$ is the
number of spin states for the particles which -- in the physical case
of electrons -- is equal to 2. (In nuclear physics one sometimes
considers $q=4$.)  We shall, however, call the particles electrons.
Similarly we write for any function $f$ depending on space and spin
variables
\begin{equation}
        \label{int}
        \int  f(\X) \, d\X= \sum_{\sigma=1}^q \int_{\R^3}\, f(\x,\sigma)\, d\x,
\end{equation}
i.e., $\int d\X$ indicates integration over the whole space and
summation over all spin indices. This allows us to write the density
matrix $\gamma$ as an operator on the Hilbert space of spinors $\psi$
for which $\int |\psi(\X)|^2\, d\X<\infty$. Its integral kernel is
$\gamma(\X,\X')$.

The Schr\"odinger Hamiltonian we wish to consider is 
\begin{equation} \label{ham}
H=\sum_{i=1}^N \left( -  \frac{\hbar^2}{2m}\nabla_i^2 -e^2 V_c(\x_i)\right)
+ e^2R 
\end{equation}
where 
\begin{equation}
V_c(\x)  = \sum_{j=1}^K \frac{Z_j}{|\x-\RR_j|}
\end{equation}
is the Coulomb potential of $K\geq 1$ fixed nuclei acting on the $N$
electrons. The $j^{\rm th}$ nucleus has charge $+Z_je>0 $ and is
located at some fixed point ${\mathbf R}_j \in \R^3$.  We define the
total nuclear charge by $Z\equiv \sum_{j=1}^K  Z_j$.  The
electron-electron repulsion $R$ is given by
\begin{equation} 
R= \sum_{1\leq i < j \leq N} |\x_i-\x_j|^{-1} \,.
\end{equation}

If one is interested in minimizing over the nuclear positions
${\mathbf R}_j$, one also has to take into account the
nucleus-nucleus repulsion $e^2 U$, of course, which is given by
\begin{equation} 
U=\sum_{1\leq i < j \leq K}Z_iZ_j |\RR_i
-\RR_j|^{-1}.
\end{equation}
Since we will not be concerned with this question but rather consider
the nuclei to be fixed, we will not take this term into consideration
here.

\subsection{Hartree-Fock Exchange Energy}

The best known density-matrix-functional associated with
\eqref{ham} is the {\it Hartree-Fock} functional
\begin{equation}
        \label{hfeqn}
        \E^{\rm HF}(\gamma) = \frac{\hbar^2}{2m} \tr (-\nabla^2 \gamma) -e^2
        \int_{\R^3} V_c(\x) \rho_\gamma(\x) d\x + e^2
        D(\rho_\gamma,\rho_\gamma) -e^2 X(\gamma)\,,
\end{equation}
where $\rho_\gamma(\x) =
\sum_{\sigma=1}^q\gamma(\X,\X)  =
\sum_{\sigma=1}^q\gamma(\x,\sigma,\x,\sigma)$
is the particle density, 
\begin{equation}
\label{D}
D(\rho,\mu) = \frac{1}{2}\int_{\R^3}\int_{\R^3}
\frac{\rho(\x)\mu(\y)}{|\x-\y|} d\x d\y,
\end{equation}
and where the exchange term is (note the sign in \eqref{hfeqn}) 
\begin{equation}
        \label{X}
        X(\gamma)= \frac12\int \int \, \frac{|\gamma(\X,\X')|^2}{|\x-\y|} \,  d\X \, d\X' \  .
\end{equation}

As is well known, this functional $\E^{\rm HF}$ is the expectation value of $H$
      in 
a determinantal wavefunction $\Psi $ made of orthonormal functions $\phi_i$
\begin{equation}
\Psi (\X_1, \, \X_2,\ \dots \, \X_N) = (N!)^{-1/2}\, {\rm det} \phi_i(\X_j)|_{i,j=1}^N\ ,
\end{equation} 
in which case 
\begin{equation} \label{hfgamma}
\gamma(\X,\X') = \sum_{i=1}^N \phi_i(\X) \phi_i(\X')^* \ .
\end{equation}

It is also well known that {\it any} one-body density matrix $\gamma$
for fermions always has two properties (in addition to the obvious
requirement of self-adjointness, i.e., $\gamma(\X,\X') =
\gamma(\X',\X)^*$) which are necessary and sufficient to ensure that it
comes from a normalized $N$-body state  satisfying the Pauli exclusion
principle, see e.g., \cite{Lieb1983D,Lieb1985}:
\begin{equation}
        \label{props}
        0\leq \gamma
\leq 1 \ {\rm as \ an \ operator} \quad {\rm and} \quad \tr \gamma =N,
\end{equation}
where $\tr$ denotes the trace $= \int d\X\,\gamma(\X,\X) =$ sum of the
eigenvalues of $\gamma$.  A simple consequence of \eqref{props} is that the spin-summed density matrix 
$(\tr_\sigma \gamma ) (\x, \x') = \sum_\sigma \gamma (\x, \sigma, \x' \sigma) $,
which acts on functions of space alone, 
satisfies
\begin{equation}
        \label{propssum}
        0\leq \tr_\sigma \gamma
\leq q \ {\rm as \ an \ operator} \quad {\rm and} \quad \tr (\tr_\sigma \gamma) =N.
\end{equation}

The HF $\gamma$ in \eqref{hfgamma} has $N$
eigenvalues equal to 1, and the rest equal to 0, but one could ignore
this feature and  apply \eqref{hfeqn} to any $\gamma$
satisfying \eqref{props}.  If we do this, then we can define the HF
energy (for all $N\geq 0$) by 
\begin{equation}
        \label{hfenergy}
        E^{\rm HF}(N) = \inf_\gamma \{\E^{\rm HF}(\gamma) :
\ 0\leq \gamma \leq1, \tr \gamma =N\} \ .
\end{equation}
(We say `infimum' in \eqref{hfenergy} instead of `minimum' because
there may be no actual minimizer -- as occurs when $N \gg Z = \sum_j
Z_j$.) A HF energy minimizer does exist when $N<
Z+1$, at least, and possibly for larger $N$'s as well
\cite{LiebSimon1974,LiebSimon1977T}.

It is a fact \cite{Lieb1981} (see also \cite{Bach1992})
that 
$ E^{\rm HF}(N) $ is the infimum over all $\gamma$'s of 
the determinantal 
form  \eqref{hfgamma}, i.e.,  the determinantal
functions always win the competition in \eqref{hfenergy}. 
Therefore,    $E^{\rm HF}(N) \geq E_0(N)$,
where $E_0(N)$ is the true ground
state energy of the Hamiltonian \eqref{ham}.

Thus, the HF density-matrix-functional has the advantage of providing
an upper bound to $E_0 $, but it cannot do better than HF theory.  We
know, however, that this is often not very good, numerically,
especially for dissociation energies.

Another disadvantage of $\E^{\rm HF}$ is that the energy minimizer
$\gamma^{\rm HF} $ (if there is one) may not be unique although, in
some cases, it is known to be unique (see \cite{HuberSiedentop2007}
for the Dirac-Fock equations). In fact it follows from Hund's rule
that in many cases the spatial part of the wave function has a
non-zero angular momentum and cannot, therefore, be spherically
symmetric.

A third point to note is that in HF theory the electron Coulomb
repulsion is modeled by $D(\rho_\gamma, \rho_\gamma) - X(\gamma) $.
This energy really should be $\int_{\R^3}\int_{\R^3} |\x-\y|^{-1} \,
\rho^{(2)}(\x,\y)d\x d\y$, however, where $\rho^{(2)}(\x,\y)$ is the
two-particle density, i.e., the spin summed diagonal part of the
\textit{two-particle} density matrix. In effect, one is replacing
$\rho^{(2)}(\x,\y)$ by $G^{(2)}(\x,\y) =
\frac12\rho_\gamma(\x)\rho_\gamma(\y) - \frac12
\sum_{\sigma,\sigma'=1}^q|\gamma(\X,\X')|^2$.  It is not possible for
this $G^{(2)}$ to be the two-body density of any state because that
would require that $\int_{\R^3} G^{(2)}(\x,\y) d\y = \frac{N-1}{2}
\rho(\x)$.  This condition fails unless the state is a HF state
(because even the total integral is wrong, namely, $\int\!\!\int
G^{(2)} d\x d\y > N(N-1)/2$ unless we have a HF state).

\subsection{M\"uller's Square-Root Exchange-Correlation Energy}
\label{1B}

There is an alternative to $\E^{\rm HF}(\gamma)$, which we will call
$\E^{\rm M} (\gamma)$ (M\"uller \cite{Muller1984}). It replaces the operator
$\gamma$ in $X(\gamma) $ by $\gamma^{1/2}$.  This means the {\it
operator} square root (note that $\gamma$ is self-adjoint and positive
as an operator, so the square root is well defined).  Thus,
$\gamma(\X,\X') = \int d\X''\, \gamma^{1/2}(\X,\X'') \gamma^{1/2}(\X'',\X')$. 
In
terms of spectral representations, with eigenvalues $\lambda_i $ and
orthonormal eigenfunctions $\phi_i$ (the `natural orbitals'),
\begin{equation}
        \label{exp}
        \gamma(\X,\X') = \sum_{i=1}^\infty \lambda_i
        \, \phi_i(\X) \phi_i(\X')^* \quad {\rm and} \quad \gamma^{1/2} (\X,\X')
        = \sum_{i=1}^\infty \lambda_i^{1/2} \phi_i(\X) \phi_i(\X')^* .
\end{equation}
There is no  simple formula for the calculation of $\gamma^{1/2}(\X,\X')$
in terms of $\gamma(\X,\X')$, unfortunately, but there is an integral representation,
which we shall use later.

Thus, 
\begin{equation} \label{mullerfunct}
        \E^{\rm M}(\gamma) = \frac{\hbar^2}{2m} \tr (-\nabla^2 \gamma) -e^2
        \int_{\R^3} V_c(\x) \rho_\gamma(\x) d\x +e^2D(\rho_\gamma,
        \rho_\gamma) -e^2 X(\gamma^{1/2}) \ ,
\end{equation}
and
\begin{equation}
        \label{sqrtenergy}
        E^{\rm M}(N) = \inf_\gamma \{\E^{\rm M}(\gamma)\, : \,
        0\leq \gamma \leq1, \tr \gamma =N\} \ .
\end{equation}

The functional $\E^{\rm M}(\gamma) $ was introduced by M\"uller
\cite{Muller1984} and was rederived by other methods by Buijse and
Baerends \cite{BuijseBaerends2002}.  A similar functional was
introduced by Goedecker and Umrigar \cite{GoedeckerUmrigar1998}, the
chief difference being that \cite{GoedeckerUmrigar1998} attempts to
remove an electron `self-energy' by omitting certain diagonal terms
that arise when \eqref{sqrtenergy} is explicitly written out using the
expansion of $\gamma$ into its orbitals \eqref{exp}.
In particular, quite analogous to
density functional theory, explicit corrections terms have been added
to correct the overestimate of binding energies using M\"uller's functional
(Gritsenko et al. \cite{Gritsenkoetal2005}).

{}From now on we will use atomic units, i.e., $\hbar=m=e=1$.  To get
some idea of the magnitudes involved we can look at hydrogen.
Numerical computations \cite[Figure 6]{Gritsenkoetal2005} and
\cite[Figure 3.1]{Helbig2006} suggest that $E^{\rm M}(1)\approx
-0.525$. This is to be compared with the true energy, $-0.5$.

It might be wondered how M\"uller's exchange energy compares to the
old Dirac $-\int \rho_\gamma(\x)^{4/3} d\x$.  As remarked after Lemma
\ref{xhardy}, and as found earlier by Cioslowski and Pernal
\cite{CioslowskiPernal1999}, $ X(\gamma^{1/2}) $ can not be bounded by
$C\int \rho_\gamma(\x)^{4/3} d\x$ for any $C$.

M\"uller \cite{Muller1984} also considered using $\gamma^{p}(\X,\X')
\gamma^{1-p}(\X',\X)$ for some $0<p<1$ in place of
$|\gamma^{1/2}(\X,\X')|^2=\gamma^{1/2}(\X,\X') \gamma^{1/2}(\X',\X)$, which
satisfies the integral condition, but he decided to take $p=1/2$
because this yields the smallest value of $X$, and hence the largest
energy.  (The proof is analogous to $a^p b^{1-p} + a^{1-p}b^p \geq
2\sqrt{ab}$ for positive numbers $a,b$.)

M\"uller's functional \eqref{mullerfunct} has several advantages, the
first of which is

{\bf A.1.} The quantity that effectively replaces $\rho^{(2)}(\x,\y)$
in the functional is now
$$\frac12\rho_\gamma(\x)\rho_\gamma(\y) -\frac12\sum_{\sigma,\sigma'=1}^q
|\gamma^{1/2}(\x,\sigma,\  \y, \sigma')|^2 , $$ 
and this satisfies the correct integral condition
$$ \frac{1}{2}\int \left[\rho_\gamma(\x)\rho_\gamma (\y) -
        \sum_{\sigma,\sigma'=1}^q\gamma^{1/2}(\X,\X') \gamma^{1/2}(\X',\X) \right]
        d\y= \frac{N-1}{2} \rho_\gamma (\x).
$$ On the other hand, $\rho_\gamma(\x)\rho_\gamma(\y) -
\sum_{\sigma,\sigma'=1}^q|\gamma^{1/2}(\X,\X')|^2$ is not necessarily
positive as a \textit{function} of $\x,\y$, whereas the HF choice
$\rho_\gamma(\x)\rho_\gamma(\y) - \sum_{\sigma,\sigma'=1}^q|\gamma(\X,\X')
|^2\geq 0$ (which is true for any positive semi-definite operator).
This non-positivity is a source of some annoyance. In particular, it prevents the
application of a standard method  \cite{Lieb1984} for
proving a bound on the maximum $N$.

{\bf A.2.} 
A special choice of $\gamma$ is a HF type of $\gamma$, namely one in
which all the $\lambda_i$ are 0 or 1. In this special case $\gamma^{1/2} =
\gamma$ and the value of the M\"uller energy equals the HF energy.
Thus, the M\"uller functional is a
generalization of the HF functional, and its
energy satisfies $E^{\rm M}(N)\leq E^{\rm HF}(N)$
(because, as we
remarked above, the minimizers for the HF problem always have this
projection property).

Later, we shall
propose that the quantity $\eh(N) = E^{\rm M} (N) +N/8$ should  be
interpreted as the binding energy; it is not obvious that $\eh(N)$
satisfies such an inequality, however. Indeed, it does not, in
general, as the hydrogen example shows ($ -0.525 + 1/8 > -0.5$).

{\bf A.3.}  The original M\"uller functional seems to give good
numerical results when few electrons are involved. Moreover, $E^{\rm
M}(N)$ appears to satisfy $E^{\rm M}(N) \leq E_0(N)$ for all electron
numbers $N$, i.e., it is always a lower bound.  We shall {\it prove}
this inequality when $N=2$ in the last section. (Numerical accuracy of
larger electron numbers seem to require appropriately modified
functionals.  We refer the reader interested on numerical results and
improved density matrix functionals to the papers of Buijse and
Baerends \cite{BuijseBaerends2002}, Staroverov and Scuseria
\cite{StaroverovScuseria2002}, Herbert and Harriman
\cite{HerbertHarriman2003}, Gritsenko et al. \cite{Gritsenkoetal2005},
Poater et al. \cite{Poateretal2005}, Lathiotakis et al.
\cite{Lathiotakisetal2005}, and Helbig \cite{Helbig2006}.)  Since we
are primarily interested in the structure of the underlying theory
rather than numerical results, we concentrate on the unmodified
original M\"uller functional despite the above mentioned numerical
deficiency for large electron number. The M\"uller functional can be
viewed as a prototype of density matrix functionals with simple
structures, but which are potentially useable as the basis of more
elaborate functionals, e.g.,
\cite{GoedeckerUmrigar1998,CsanyiArias2000,Csanyietal2002,Gritsenkoetal2005}.


\subsection{Convexity and Some of  its  Uses}

A key observation about $\E^{\rm M}(\gamma)$ is that it is a {\it convex
       functional} of $\gamma$. This means that for all
$0<\lambda<1 $ and density  matrices $\gamma_1, \gamma_2$ (not
necessarily with the same trace and not necessarily satisfying
$\gamma \leq 1$)
\begin{equation}  \label{convexity}
\E^{\rm M}( \lambda \gamma_1 + (1-\lambda)\gamma_2) \leq 
\lambda \E^{\rm M} ( \gamma_1 )+ (1-\lambda)\E^{\rm M}(\gamma_2) \ .
\end{equation}
(Note that the convex combination $\lambda \gamma_1 +
(1-\lambda)\gamma_2$ satisfies the conditions in \eqref{sqrtenergy} if
$\gamma_1$ and $\gamma_2$ both satisfy the conditions.)  The convexity
is a bit surprising, given the minus sign in the exchange term of
$\E^{\rm M}$, and it will lead to several important theorems.  One is that
the electron {\it density} $\rho_\gamma(\x)$ of the minimizer (if
there is one) is the same for all minimizers with the same $N$, and
hence that the density of an atom is always spherically symmetric.
This contrasts sharply with   HF theory, whose functional
\eqref{hfeqn} is {\it not} convex, and it can contradict  the original
Schr\"odinger theory (since an atom can have a nonzero  angular momentum
in its ground state).  Also, the Dirac estimate for the exchange energy,
$-\int \rho^{4/3}$ is not convex; it is concave, in fact!

Some writers \cite{Grossetal1991} regard the retention of symmetry as a
desirable property for an approximate theory; one speaks of the
``symmetry dilemma'' of HF theory (which means that while symmetry
restriction of HF orbitals improves the overall symmetry it raises the
minimum energy). M\"uller theory has no symmetry dilemma!

{}From another perspective the sphericity of an atom might be seen as a
drawback since real atoms sometimes have a non-zero angular momentum,
and such states are not spherically symmetric.  Sphericity is not a
drawback, in fact, since density-matrix-functional theory deals with
density-matrices obtained from {\it all} $N-$particle states,
including mixed ones (because the only restriction we impose is that
the eigenvalues of $\gamma$ lie between 0 and 1, and this condition
precisely defines the set of $\gamma$ obtained from the set of mixed
states, not the set of pure states).  In the case of atoms there is
always a mixed state with spherical symmetry, namely the projection
onto all the ground states, divided by the degeneracy.  This is the
state that one sees (in principle) when looking at an atom at zero
temperature (L\"uders' projection postulate \cite{Luders1951}).

A second consequence of convexity is that the energy $E^{\rm M}(N) $ is
always a convex function of $N$, as it is in Thomas-Fermi theory, for
example  \cite{LiebSimon1977,Lieb1981}. This means that as we
add one electron at a time to our molecule, the (differential) binding
energy steadily decreases.  Such a property is {\it not} known to hold
for the true    Schr\"odinger energy $E_0(N)$.

The convexity of $\E^{\rm M}(\gamma)$   is not at all obvious. All the terms
except $-X(\gamma^{1/2})$ are clearly convex. In fact, the term
$D(\rho_\gamma, \rho_\gamma)$ is {\it strictly} convex as a function
of the density $\rho_\gamma(\x)$ (strict inequality in
\eqref{convexity} when $\rho_{\gamma_1}\neq \rho_{\gamma_2}$) since the Coulomb
kernel $|\x-\y|^{-1} $ is positive definite. It is this strict
convexity that implies the uniqueness of $\rho_\gamma(\x)$ when there
is a minimizer.

To show convexity of $\E(\gamma)$, therefore, we have to show
concavity (like \eqref{convexity} but with the inequality reversed) of
the functional $X(\gamma^{1/2})$.  First, we write $|\x-\y|^{-1} =
\int_\Lambda B_\lambda(\x)^*B_\lambda(\y) d\lambda $ where
$\lambda $ is in some parameter-space $\Lambda$.
There are many ways to construct such a decomposition.
One way is due to Fefferman and de la Llave \cite{FeffermandelaLlave1986},
which we shall use in the sequel, in which the functions $B_\lambda$
are all characteristic functions of balls in $\R^3$ and $\lambda$
parametrizes their radii and centers. Another way is $|\x-\y|^{-1} =
C\int_{\R^3} |\x-\z|^{-2} |\y-\z|^{-2} d\z$.  Anyway, it suffices now
to prove that $\int d\X\, d\X'\, \gamma^{1/2}(\X, \X') B(\x)^*
\gamma^{1/2}(\X', \X) B(\y)$ is concave in $\gamma$, for any fixed
function $B(\x)$. We can write this in abstract operator form as $\tr
\gamma^{1/2} B^\dagger \gamma^{1/2} B$. The concavity of such functions of
$\gamma$ was proved  by Wigner and Yanase
\cite{WignerYanase1964} in connection with a study of entropy.

Convexity also holds for M\"uller's general $p$ functional, which we 
mentioned earlier.  It uses $\gamma^{p}(\X,\X')\gamma^{1-p}(\X',\X)$ in the 
exchange term. The fact  that  $\tr
\gamma^{p} B^\dagger \gamma^{1-p} B$ is concave for all $0<p<1$
      was proved in \cite{Lieb1973C} and plays a role
in quantum information theory \cite{NielsenChuang2000}.

Another important use of the convexity of $\E^{\rm M}(\gamma)$ is to significantly
simplify the question of the spin dependence of $\gamma(\x,\sigma, \y,
\sigma')$.  For concreteness, let us assume the usual case of two spin
states ($q=2$), but the conclusion holds for any $q$. In the HF
problem it is not obvious how $\gamma$ should depend on $\sigma,
\sigma'$ and usually one makes some standard a-priori assumption, such
as that $\gamma^{\rm HF}(\x,\sigma, \y, \sigma') =
\gamma_{\uparrow,\uparrow} (\x,\y, )  \delta_{\sigma, \uparrow}
\delta_{\sigma', \uparrow} + \gamma_{\downarrow,\downarrow} (\x,\y, )
      \delta_{\sigma, \downarrow} \delta_{\sigma', \downarrow} $.  In
the M\"uller case this problem does not arise. Note that the
functional $\E^{\rm M}$ is invariant under simultaneous rotation of $\sigma
$ and $\sigma'$ in spin-space.  (This means that we regard $\gamma$ as
a $2\times 2$ matrix whose elements are function of $\x, \y$. The spin
rotation is then just a $2\times2$ unitary transformation of this
matrix.)  If we take any $\gamma(\x,\sigma, \y, \sigma')$ and average
it over all such simultaneous rotations we will obtain a new
$\widetilde\gamma$ whose energy $\E^{\rm M}(\widetilde\gamma)$ is at
least as low as that of the original $\gamma$ (by convexity). But
$\widetilde\gamma$ is clearly spin-space rotation invariant, which
means it must have the form \be
\label{gammaform} \widetilde\gamma (\x,\sigma, \y, \sigma') =
\frac12 \widehat\gamma (\x, \y) \otimes \I
\end{equation}
where $\I$ is the $2\times 2$ identity matrix. The subsidiary  conditions 
become
\be  \label{newcondition}
\tr \widehat\gamma \equiv \int \widehat\gamma (\x,\x) d\x =N
\qquad\qquad  {\rm and } \qquad\qquad 0\leq \widehat\gamma \leq 2 \ .
\end{equation}
The change from 1 to 2 in \eqref{newcondition} is to be noted. 
Often $\widehat\gamma $ is called the {\it spin-summed density matrix}.

The conclusion is that to get the correct minimum energy one can
always restrict attention to the simpler, spin-independent
$\widehat\gamma$, but with the revised conditions
\eqref{newcondition}.  This is  a significant simplification relative to HF theory.
In much of the sequel we utilize the formal
notation $\X$ instead of $\x$, but the reader should keep in mind that
one can always assume that $\gamma$ has the form \eqref{gammaform} and
all spin summations become trivial.

A question will arise: Although it is possible to choose $\gamma $ in
the form \eqref{gammaform}, are there other possibilities?  They will
certainly exist if $\widehat\gamma $ is not unique, (but we conjecture
that it is unique since its density is unique, as we said).  Even if
$\widehat\gamma $ is unique we still might have other possibilities,
however, when $N$ is small.  For example, we could take $\gamma (\X,
\X') = \widehat\gamma (\x,\y) \times \delta_{\sigma, \uparrow}
\delta_{\sigma', \uparrow} $, but this density matrix is bounded above
by 1 only if $\widehat\gamma \leq 1$ (not $\leq 2)$.  This situation
can arise if $N$ is small, but we expect that it does not arise when
$N\geq 1$. In any case, we show that, for large $N$ and $Z$,
$\widehat\gamma $ has at least one maximal eigenvalue, namely 2 (see
Prop.~\ref{propq}).

In short, it is likely that whatever the M\"uller functional has to
say about the energy, it probably has little to say, reliably, about
the spin of the ground state. Unlike HF theory, we do not have to
worry about spin here. This does not mean that HF theory is
necessarily better as concerns spin. Sometimes it is
\cite{Bachetal1994}, 
and sometimes it is not \cite{Bachetal2006}.

In the atomic case $\E^{\rm M}(\gamma)$ is also rotationally invariant and
we can apply the same logic used above for the spin to the
simultaneous rotation of $\x, \y$ in $\widehat\gamma (\x,\y)$. The
conclusion is that we may assume the following computationally useful
representation:
\begin{equation}
\widehat\gamma (\x,\y)   = \sum_{\ell=0}^\infty \sum_{m=-\ell}^\ell 
\gamma_\ell(r,r') 
Y_{\ell, m}(\theta_\x)\, Y_{\ell, m}^*(\theta_\y) 
=\frac{2\ell +1}{4\pi} \sum_{\ell=0}^\infty  \gamma_\ell(r ,  r')  P_\ell (\cos \Theta) ,
\end{equation}
where $r=|\x|, r'=|\y|$.  The $Y_{\ell, m}$ are normalized spherical
harmonics, $\theta_\x$ is the angle of the vector $\x$, etc., $P_\ell
$ is the $\ell^{\rm th}$ Legendre polynomial and $\Theta $ is the
angle between $\x$ and $\y$.  Another way to say this is that we can
assume that the eigenfunctions of $\widehat\gamma (\x,\y) $ are radial
functions times spherical harmonics $Y_{\ell, m}$ and that the allowed $m$
values occur with equal weight. 
This observation can simplify
numerical computations.

Any other symmetry can be treated in a similar way. For example, in
the case of a solid there is translation invariance of the lattice of
nuclei. By wrapping a large, finite piece of the lattice on a torus
(periodic boundary conditions) we have a finite system with
translation invariance and we can conclude, as above, that we can
assume that $\widehat\gamma (\x,\y) $ is also translation invariant,
which means that $\widehat\gamma (\x,\y) $, viewed as a function of
$\x+\y$ and $\x-\y$ is periodic in the variable $\x+\y$.

One obvious symmetry is complex conjugation ($i\to -i$ ) in the absence of a 
magnetic field. Convexity implies that in the spin-independent formulation
any minimizing $\gamma $ must be real, as shown in Proposition \ref{realmin}
of Section \ref{minprop}.

\subsection{The M\"uller Equations\label{subsec:Mueller}}

If the M\"uller functional has a minimizing $\gamma$ (with $\tr \gamma
=N$) then this $\gamma$ satisfies an Euler equation. A minimizer does
exist if $N\leq Z$ as we show in Theorem \ref{fulltrace}.  It is not
altogether a trivial matter to write down an equation satisfied by a
minimizing $\gamma$.  Conversely, one can ask whether a $\gamma $ that
satisfies this equation is necessarily a minimizer. We partly answer
these questions in several ways.

\quad 1. Suppose that $\gamma$ satisfies $\tr \gamma =N $ and that
$\gamma$ minimizes $\E^{\rm M}(\gamma)$, i.e., $\E^{\rm M}(\gamma) =
E^{\rm M}(N)$. Then
we conclude (by definition of the minimum) that
\begin{equation} \label{min1}
\E^{\rm M}((1-t)\g + t\g') \geq \E^{\rm M}(\g) 
\end{equation}
for all admissible $\g'$ with $\tr \gamma' =N $ and for all $0\leq t
\leq 1$.  Conversely, if $\tr \gamma =N $ and if \eqref{min1} is true
for all such $\g'$ and for {\it some} $0<t\leq 1$ (with $t$ possibly
depending on $\g'$) then $\g$ is a minimizer. Alternatively, it
suffices to require that for all such $\g'$
\begin{equation} \label{min2}
\frac{d}{dt} \E^{\rm M}((1-t)\g + t\g') |_{t=0 } = \lim_{t \downarrow 0} \frac1t \left[
\E^{\rm M}((1-t)\g + t\g') - \E^{\rm M}(\g)\right] \geq 0\, .
\end{equation}
To see that $\g$ is a minimizer we exploit the convexity of the
functional $\E^{\rm M}$, which implies that $\E^{\rm M}((1-t)\g + t\g') \leq
(1-t)\E^{\rm M}(\g) +t\E^{\rm M}(\g'),$ and hence, from \eqref{min1} or
\eqref{min2}, that $\E^{\rm M}(\g) \leq \E^{\rm M}(\g')$.  (Note that the convexity
also implies that $\E^{\rm M}((1-t)\g + t\g')$ is a convex function of $t$ in
the interval $[0,1]$, which, in turn, implies that the right
derivative defined in \eqref{min2} always exists.)

To summarize, we say that {\it the equation defining a minimizer is
    \eqref{min2}} (for all $\g'$).  To make this more explicit we have
to compute the derivative in \eqref{min2}.

\quad 2. The variational equations are most conveniently written down
in terms of $\gamma^{1/2}$, the square root of a minimizer. In
Proposition~\ref{lagrange2}, we will show that $\gamma^{1/2}(\x,\y)$
satisfies the following variational equation. Let $\varphi_\gamma$
denote the effective potential $\varphi_\gamma(\x)= V_c(\x) - \int
\rho_\gamma(\y) |\x-\y|^{-1} d\y$, where $\rho_\gamma(\x)=\sum_\sigma
\gamma(\X,\X)=\sum_\sigma \int |\gamma^{1/2}(\X,\X')|^2 d\X'$ denotes
the particle density. Then
\begin{equation}\label{vareq}
\left(-\half \nabla_\x^2 - \half \nabla_\y^2 - \varphi_\gamma(\x) - \varphi_\gamma(\y) 
- \frac 1{|\x-\y|} - 2\mu\right) \gamma^{1/2}(\X,\X') = 
\sum_i 2 e_i \psi_i(\X) \psi_i(\X')^*
\end{equation}
where $\mu\leq -1/8$, $e_i\leq 0$ and $\psi_i(\X)$ is an eigenfunction
of $\gamma^{1/2}$ with eigenvalue $1$, i.e., $\int
\gamma^{1/2}(\X,\X') \psi_i(\X') d\X' = \psi_i(\X)$ for all $i$. Note
that the number of $\psi_i$'s corresponding to eigenvalue $1$ is
necessarily less than $N$.

Conversely, is it true that any $\gamma^{1/2}$ satisfying $0\leq
\gamma^{1/2} \leq 1$ (as an operator) and $\tr
\left(\gamma^{1/2}\right)^2 = \tr \gamma= N$ which is a solution to
\eqref{vareq} under the constraints mentioned above, is a minimizer of
$\E^{\rm M}(\gamma)$?  Unfortunately, we can answer this question
affirmatively only if we know that the density $\rho_\gamma(\x)$ does
not vanish on a set of positive measure. Presumably such a vanishing
does not occur, but we do not know how to prove this and leave it as an
open problem.

\quad 3. As a practical matter it is the fact that $\g$ satisfies
\eqref{vareq} that is important because it gives us equations for the
orbitals of $\g$.  A minimizer $\g$ can be expanded in natural
orbitals $\psi_j(\X)$ as
$$
\gamma(\X,\X')=\sum_j\lambda_j\psi_j(\X)\psi_j(\X')^*
$$
with corresponding occupation numbers (eigenvalues)
$0<\lambda_j\leq 1$. Then $\gamma^{1/2}(\X,\X')=\sum_j
\lambda_j^{1/2}\psi_j(\X)\psi_j(\X')^*$.  Multiplying (\ref{vareq}) by
$\psi_i(\X')$ and integrating over $\X'$ yields an eigenvalue equation
for the $\psi_i(\X)$, namely
\begin{equation}
    \left[\left(-\tfrac12\nabla^2- \phi_\gamma\right)\gamma^{1/2} 
+ \gamma^{1/2}\left(-\tfrac12\nabla^2-\phi_\gamma\right)\right] 
|\psi_i\rangle - \left(Z_\gamma+2\mu\lambda_i^{1/2}\right)|\psi_i\rangle
\\= 2 e_i|\psi_i\rangle\,.
\end{equation}
Here, $Z_\gamma$ is the operator with integral kernel
\begin{equation} \label{zgamma}
Z_\gamma(\X,\X') = 
\gamma^{1/2}(\X,\X')|\x-\y|^{-1}\,.
\end{equation}
Taking the product with $\langle
\psi_j|$, this implies, in particular, that
\begin{equation}\label{me}
\langle \psi_j| -\tfrac 12 \nabla^2 - \varphi_\gamma | \psi_i\rangle -
\frac 1{\sqrt{\lambda_i}+\sqrt{\lambda_j}} \langle \psi_j| Z_\gamma
|\psi_i\rangle = \left( \mu+ e_i\right)  \delta_{ij}.
\end{equation}
(See also Pernal \cite{Pernal2005} who derived -- although merely on a
formal level -- similar equations for more general functionals).

\quad 4. We shall show that $\gamma $ has no zero eigenvalues unless
the density $\rho_{\gamma}(\x)$ vanishes identically on a set $\Omega$
of positive measure.  We do not expect such a set to exist but we do
not know how to exclude this possibility.  Any non-zero, square
integrable function that vanishes identically outside $\Omega$ is a
zero eigenvalue eigenfunction of $\gamma$. In any case, there are no
other zero eigenvalue eigenfunctions!

Hence the orbitals $\psi_j(\X)$ form a complete set in
$L^2(\R^3\setminus \Omega)$. Formally, we can thus rewrite
Eq.~(\ref{me}) as an eigenvalue equation for a linear operator
$H_\gamma$ on $L^2(\R^3\setminus\Omega)$. Let 
\begin{equation}
    \label{eq:h1}
H_\gamma = -\tfrac12 \nabla^2 -\varphi_\gamma - \mathfrak X_\gamma\,,
\end{equation}
where $\mathfrak X_\gamma$ is the nonlocal exchange operator with matrix
elements $\langle \psi_i |\mathfrak X_\gamma|\psi_j\rangle =
(\sqrt{\lambda_i}+\sqrt{\lambda_j})^{-1} \langle \psi_i |
Z_\g|\psi_j\rangle$. Alternatively, one can write 
\begin{equation}
    \label{defxg}
\mathfrak X_\gamma = \frac 1\pi \int_0 ^\infty \frac 1{\gamma+s} Z_\g \frac
1{\g+s} \sqrt{s} \,ds\,.
\end{equation}
The variational equations are then
\begin{equation} \label{equation}
H_\gamma  |\psi_j\rangle = \mu| \psi_j\rangle 
\end{equation}
for all $j$ with $0<\lambda_j<1$, where $\mu\leq -1/8$ is the chemical
potential.
Notice that all eigenvalues in \eqref{equation} are identical, namely $\mu$.

In the subspace in which $\gamma$ has eigenvalue 1, which can only be
finite dimensional since $\tr \gamma =N$, there is an orthonormal
basis such that
\begin{equation}\label{upperequation}
H_\gamma  |\psi_j\rangle = \left(\mu +e_j\right) |\psi_j\rangle
\end{equation}
with all $e_j \leq 0$. The finite collection of numbers $\mu + e_j$
constitutes all the eigenvalues of $H_\gamma $ that are less than
$\mu$.

The reason we say that \eqref{equation} and \eqref{upperequation} are
formal is that the operator $H_\gamma$ is only formally defined by
\eqref{eq:h1}.  Both $\nabla^2$ and $\mathfrak X_\gamma$ are unbounded
operators.  Their sum is defined as a quadratic form (i.e.,
expectation values) but this form does not uniquely define the
operator sum. If we knew that there are no zero eigenvalues then the
set $\Omega $ would be empty and $\nabla^2$ would be defined as the
usual Laplacian on $\R^3$, but if $\R^3 \setminus \Omega$ has a
boundary there are many extensions of $\nabla^2$ with different boundary
conditions, and this prevents the precise specification of
\eqref{equation} and \eqref{upperequation}.  There is no problem with
the matrix elements in \eqref{me}, however, since the $\psi_i$ vanish
on the boundary of $\R^3 \setminus \Omega$.

On the other hand \eqref{vareq}, which is an equation for the {\it function}
$\gamma^{1/2} (\X,\X')$, is true on the whole space. It is not
necessary to impose any boundary conditions and $\nabla^2$ is just the
usual Laplacian -- whether or not the set $\Omega$ is empty.

Surely $\Omega $ is empty, in fact,  and the practical quantum
chemist can freely use
\eqref{equation} and \eqref{upperequation}.

\subsection{Other Considerations about the M\"uller Functional}

Let us conclude this introduction with a list of other significant
questions about $ E^{\rm M}(N)$ and with  statements about
what we can prove rigorously.

{\bf Q1.}  If there are no nuclei at all ($K=0$), and if we try to
minimize $\E^{\rm M}(\gamma)$ (with $\tr \gamma =N$, however) it is clear
that there will be no energy minimizing $\gamma$. There will, of
course, be a minimizing sequence (i.e., a sequence $\gamma_n$, $
n=1,2,....$ such that $\E^{\rm M}(\gamma_n) \to E^{\rm M}(N)$ as $n\to
\infty$. Such a sequence will tend to `spread out' and get smaller and
smaller as it spreads (always with $\tr \gamma_n =N$).  What, then, is
$ E^{\rm M}(N)$? We prove that it is \textit{exactly} given by 
\begin{equation}
        E^{\rm M}(N) = -N/8 \qquad\qquad {\rm when \ all \ }Z_j=0.
\end{equation} 
(If the units are included the energy is $-(me^4 /8\hbar^2)N$.)  A
similar calculation in the context of the homogeneous electron gas was
done by Cioslowski and Pernal \cite{CioslowskiPernal1999}.

This situation is reminiscent of Thomas-Fermi-Dirac theory
\cite{Lieb1981} where, in the absence of nuclei, the energy equals
$-(const.)N$.  This negative energy comes from balancing the kinetic
energy against the negative exchange.  In such a case it is convenient
to add $+(const.) \tr \gamma$ to $\E^{\rm M}(\gamma)$ (with $(const.)
= 1/8$ in our case) in order that $E^{\rm M}(N) \equiv 0$ when there
are no nuclei.

Another way to say this is that the energy, $-1/8$, is the {\it
        self-energy} of a particle in this theory. It has no physical or
chemical meaning but we have to pay attention to it. It is the
quantity 
\begin{equation} \label{sqrthatenergy}
\eh(N) = E^{\rm M}(N) +\frac{N}{8}
\end{equation}
that might properly be regarded as the energy of $N$ electrons in the
presence of the nuclei, i.e., $-\eh(N) $ is the physical binding (or
dissociation) energy. We do not insist on this interpretation, however. On the other
hand, if we are interested in the binding energy with fixed $N$ (e.g.,
the binding energy of two atoms to form a molecule) then it makes no
difference whether we use the difference of $\eh(N) $ or $E^{\rm M}(N)$.

The motivation here is to ensure that the ground state energy of free
electrons is zero. This can be compared with the formulation in
\cite{GoedeckerUmrigar1998} in which the `self-energy' correction is obtained by
omitting certain diagonal terms in the energy (when the energy is
written in terms of the orbitals of $\gamma$).  This procedure does not
have a natural physical interpretation and, more importantly, does not
appear to give the zero energy condition for free electrons.

This consideration leads us to the functional
\begin{equation} \label{ehatfunctional}
        \Eh(\gamma) =  \E^{\rm M}(\gamma)  +\frac{1}{8}\tr \gamma
\end{equation}
and its corresponding infimum $\eh (N)$. Note that $\Eh(\gamma)$ is also
a convex functional of $\gamma$ since the new term $\tr \gamma /8$ is
linear, and hence convex. Likewise, $\eh(N) $ is a convex function of $N$.

Having added this term, and with nuclei present, $\eh(N)$ will
qualitatively look like the Thomas-Fermi energy, $E^{\rm TF}(N)$. That is,
$\widehat E^{\rm M}(0)=0$ and $\widehat E^{\rm M}(N)$ decreases
monotonically, and with non-decreasing derivative, as $N$ increases
\cite{LiebSimon1977}, \cite[Fig. 1]{Lieb1981}.  It is bounded below,
that is,
\begin{equation} \label{lowerbound-eh} 
\eh(N) \geq \eh(\infty) ,
\end{equation}
where
$\eh(\infty) $ is some finite, negative constant.  We shall {\it prove} this here. These 
features are displayed schematically in Fig. 1.

There is another feature of $E^{\rm TF}(N)$ that we believe to be true for
$\eh(N)$, but leave as an {\it open question}.  At a certain critical
value, $N_c$, of the electron number $E^{\rm TF}(N) $ stops decreasing and
becomes constant for all $N\geq N_c$.  When $N>N_c$ the excess charge
$N-N_c$ just leaks off to infinity.  In TF theory $N_c $ is the
neutrality point $Z =\sum Z_j$, but this need not be so in other
theories. In the original Schr\"odinger theory (\ref{ham}) $N_c$ is
greater than $Z$ for many atoms (since stable, negative ions exist)
but we know it is less than $2Z+1$ \cite{Lieb1984}. In the
Thomas-Fermi-Weizs\"acker theory, $N_c $ is approximately $Z+
(const.)$ \cite{BenguriaLieb1985,Lieb1981}.  We do not know
how to prove that there is a finite $N_c$ for $\eh(N)$, but we
believe there is one.


\begin{figure}[htf] 
\includegraphics[width=13cm, height=10cm]{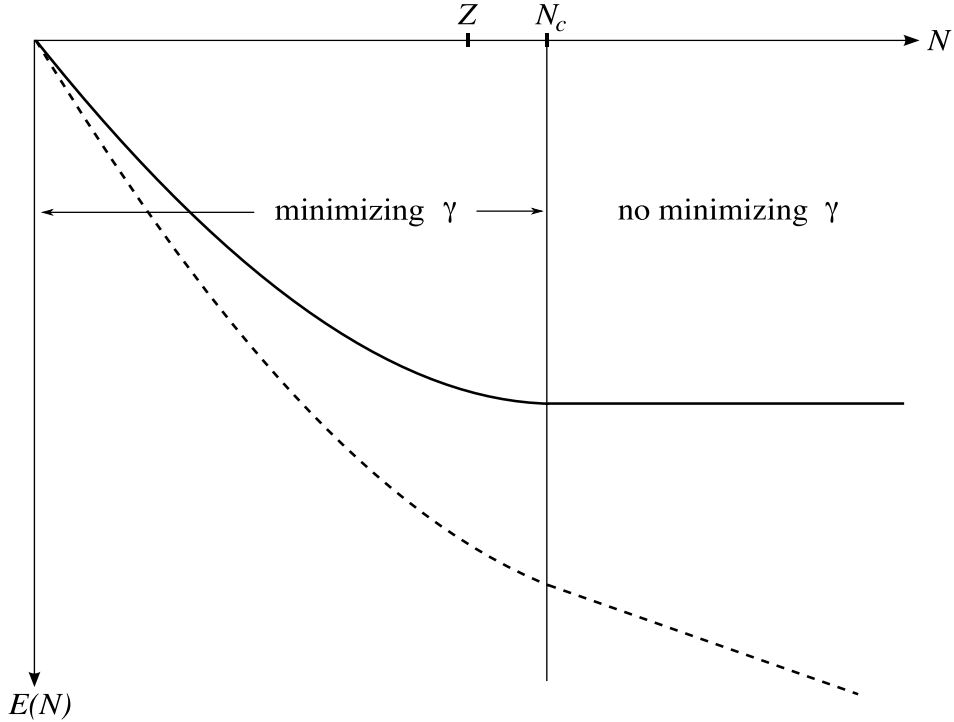}
\caption{\normalsize {Schematic diagram of the energy dependence on the particle
number $N$. The lower, dashed curve is the M\"uller energy $E^{\rm M}(N)$
and the upper, solid curve is $\eh(N) = E^{\rm M}(N) + N/8$, in which the 
`self-energy' $-N/8$ has been subtracted. Beyond the value $N_c$ each curve
is linear, whereas for $N < N_c $ each  is strictly convex and there is an
energy minimizing density matrix.}}
\label{Fig.1}
\end{figure}


{\bf Q2.} The main problem that has to be addressed is whether or not
there is a $\gamma$ that minimizes $\Eh(\gamma)$ in
\eqref{sqrtenergy}.  If $N_c <\infty$ we know that there is {\it no}
minimizer when $N>N_c$, so we obviously do not expect to prove the
existence of a minimizer for all $N$.

The way around this problem, as used in \cite{LiebSimon1977}, for
example, is to consider the {\it relaxed problem} 
\begin{equation}
         \label{relaxed}
        \eh_\leq (N) = \inf_\gamma \{\Eh(\gamma) \, : \,  0\leq \gamma \leq1, \tr \gamma \leq N\} \ .
\end{equation}
The relaxation of the number condition allows electrons to move to
infinity in case $N$ is larger than the maximal number of electrons
that can be bound.  In Proposition \ref{minz=} we show that $\eh_\leq
(N) = \eh (N)$ for all $N$.

The difference is that while the $\eh $ problem may not have a
minimizer we prove that the $\eh_\leq$ problem \eqref{relaxed} has a
minimizer for all $N$.  The proof is more complicated in several ways
than the analogous proof in TF theory \cite{LiebSimon1977,Lieb1981}.
A minimizer, which we can call $\gamma_\leq(N)$, will have some
particle number $\tr \gamma_\leq(N) \equiv N_\leq \leq N$.  It then
follows from standard arguments using convexity (and strict convexity
of $D(\rho, \rho)$) that the following is true, as displayed in Fig.
1 :

If $N_\leq < N$ then $\gamma_\leq(N) = \gamma_\leq(N_\leq) $ and 
$\eh(N) = ({\rm constant}) = \eh(N_\leq) $, i.e.,
the original problem \eqref{sqrtenergy} has no minimizer.

If $N_\leq =N$ then $\gamma_\leq(N) $ is also a minimizer for the
original problem \eqref{sqrtenergy}.  That is, the relaxed problem and
the original problem give the same minimizer and the same energy.  In
this case, $\eh(N) < \eh(N')$ for all $N'< N$.  The largest $N$ with
this property is equal to $N_c$.

It might occur to the reader that nothing said so far precludes the
possibility that $N_c = 0$, but this is not so. We prove that $N_c
\geq Z$ = total nuclear charge.

{\bf Q3.}  How many orbitals are contained in  a minimizing $\gamma$?
We shall prove that $\gamma$ has infinitely many positive eigenvalues.
This feature also holds for the full Schr\"odinger theory (Friesecke
\cite{Friesecke2003} and Lewin \cite{Lewin2004}), whereas there are
only $N$ in HF theory.  We believe that $\gamma$ has no zero
eigenvalues (in the `spin-summed' version), but cannot prove this. In
other words, we believe that the eigenfunctions belonging to the
nonzero eigenvalues span Hilbert space (they form a complete set). We
can, however, prove that the eigenfunctions of the spin-summed
$\gamma$ are a complete set on the support of $\rho_\gamma(\x)$, namely on
the set of $\x \in \R^3$ for which $\rho_\gamma(\x) >0$. Presumably,
this is the whole of $\R^3$.

This introduction is long, but we hope it serves to clarify our goals
and results, since the rest of the paper is unavoidably technical.



\subsection{Open Problems}

For the reader's convenience we give a brief summary of some of the open problems
raised by this work, some of which are discussed at various places in this paper.

\begin{enumerate}
\item What is the critical value of the total electron charge, $N_c$,
    beyond which there is no energy minimizing $\gamma$ and the energy
    $\eh(N)$ is constant? Is $N_c$ finite and can one give upper and
    lower bounds to it? In particular is $N_c>Z$, i.e., can negative
    ions exist?  (We prove $N_c \geq Z$ and we prove that $\eh(N)$ is
    bounded below, for all $N$, by a $Z$-dependent constant.)

\item Is $E^{\rm M}(N) \leq$ the true Schr\"odinger ground state
    energy? (We prove this for $N=2$.) Can anything be said, in this
    regard, about $\eh(N) = E^{\rm M}(N) +N/8 $?

\item Is the spin-summed energy minimizing $\gamma$ unique?
(We prove that all minimizers have the
same density $\rho(\x)$, however.)

\item Is the domain on which the unique $\rho(\x) >0$ equal to the
    whole of $\R^3$ (except, possibly, for sets of measure zero)?  If
    so, this would imply that the spin-summed $\gamma$ does not have a
    zero eigenvalue.

\item What are the qualitative properties of the density $\rho(\x)$?
    How does it fall off for large $|\x|$? What is its behavior near the
    nuclei?

\item In this theory do atoms bind to form molecules? (Recall that
    there is no binding in Thomas-Fermi theory \cite{LiebSimon1977}.)

\end{enumerate}


\section{The Case $Z=0$} \label{sec2}

As noted in the introduction the energy of {\it free} electrons
$E^{\rm M}(N)$ is not zero but is proportional to $N$. To be precise,
$E^{\rm M}(N) =-N/8$ (in atomic units) when there are no nuclei, and
comes about from the negative exchange energy $-X(\gamma^{1/2})$. This
negative energy could be $-\infty$ were it not for the positive
kinetic energy, which controls it and leads to a finite result. We
shall prove that the direct Coulomb repulsion term, $D(\rho_\gamma,
\rho_\gamma)$ plays {\it no role} here because it is quadratic in
$\gamma$, whereas the terms we are concerned with are homogeneous of
order $1$.  We
would get $-N/8$ even if we omitted the direct term.  Similarly, the
value $-N/8$ is independent of the number of spin states $q$.
Moreover, the assumption $\gamma\leq 1$ is {\it not} needed in the
proof.

In this section, $Z= \sum Z_j= 0$, and we are 
considering  the functional
\begin{equation} \label{z=0functional}
        \mathcal E^{\rm M}(\gamma) \equiv \tr(-\half\nabla^2\gamma) +
        D(\rho_\gamma,\rho_\gamma)-X(\gamma^{1/2})
\end{equation}
and the minimal energy $E^{\rm M}(N)$ in \eqref{sqrtenergy}.  We also
consider the relaxed energy $E^{\rm M}_{\leq}(N)$ for which, in
analogy with \eqref{relaxed}, the condition $\tr \gamma =N $ is
replaced by $\tr \gamma \leq N $.

We always assume that
$(-\nabla^2+1)^{1/2}\gamma^{1/2}\in\mathfrak S^2$, the set of
Hil\-bert-Schmidt operators, so $\tr((1-\nabla^2)\gamma) = \int 
\int d\X\, d\X'\,
\left(|\nabla\gamma^{1/2}(\X, \X')|^2 +|\gamma^{1/2}(\X,\X')|^2\right)
       <\infty$.
We use the usual notation for $L^p$-norms, namely
$$\Vert f \Vert_p = \left(\int |f(\X)|^p d\X \right)^{1/p}\ \text{and}\ 
\Vert f \Vert_\infty = \sup_\X \{|f(\X )|\}.
$$

\begin{proposition}\label{z0}
If $Z=0$, then for any $N>0$,
\begin{equation}
E^{\rm M}(N) = E^{\rm M}_\leq(N) = -N/8
\end{equation}
and there is no minimizing $\gamma$.
\end{proposition}

\begin{proof}
\underline{Lower bound}: We use the lower semi-boundedness of the hydrogenic Hamiltonian
(i.e., for an imaginary nucleus with $Z=1/2$, located at $\y$) 
\begin{equation}\label{eq:hydrogene}
              -\tfrac12\nabla^2_\x - (2|\x-\y|)^{-1} \geq -\tfrac18
\end{equation}
for all $\y\in\R^3$, together with the fact that
$D(\rho_\gamma,\rho_\gamma)\geq 0$ to get
\begin{align*}
        \mathcal E^{\rm M}(\gamma)
        & \geq \frac12\iint \left(|\nabla_\x \gamma^{1/2}(\X,\X')|^2 -
        \frac{|\gamma^{1/2}(\X,\X')|^2}{|\x-\y|}\right)\,d\X\,d\X' \\
        & \geq -\frac18 \iint |\gamma^{1/2}(\X,\X')|^2 \,d\X\,d\X' =-\frac18\tr\gamma.
\end{align*}
This proves the lower bound on $E^{\rm M}(N)$ and $E^{\rm M}_\leq (N)$.

To prove the non-existence of a minimizer we denote by $g(\x-\y)$ the
ground state of
$-\nabla_{\x}^2-|\x-\y|^{-1}$, i.e.,
\begin{equation}\label{eq:g}
g(\x-\y) = \pi^{-1/2} e^{-|\x-\y|},
\end{equation}
and note that the inequality $\leq $ in \eqref{eq:hydrogene} is strict
(i.e., it is $>$), except for multiples of the function $g(\x
-\y)$. Hence the above lower bound on $\mathcal E^{\rm M}(\gamma)$ is strict
unless $\gamma^{1/2}(\X,\X') = c_{\sigma\sigma'}(\y)g(\x-\y)$.  By
self-adjointness, $c_{\sigma\sigma'}$ has to be a constant, and since
$\gamma\in\mathfrak S^1$, the set of trace class operators,
$c_{\sigma\sigma'}=0$. But this means that there exists no minimizer.

\underline{Upper bound}: We define a trial density matrix $\gamma$ by
defining its square root:
\begin{equation}\label{eq:freetrial}
\gamma^{1/2}(\X,\X') = \chi (\x)^* g(\x-\y) \chi (\y) \, q^{-1/2}\delta_{\sigma,\sigma'}.
\end{equation}
Here, $g$ is the same as in \eqref{eq:g} and $\chi$ is a smooth
function which will be specified later. Note that this definition
makes sense, since the operator whose kernel is given on the right side
of \eqref{eq:freetrial} is non-negative. This follows from the
positivity of $\widehat{g}$, the Fourier transform of $g$, given by
\begin{equation*}
\widehat g(\p) = \frac{2^{3/2}}\pi\frac1{(1+|\p|^2)^2}.
\end{equation*}
An easy calculation shows that
\begin{equation*}
\tr(-\nabla^2_\x\gamma) = \iint \left(|\chi(\x)|^2 
|\chi(\y)|^2 (-\nabla^2_\x g(\x-\y))
g(\x-\y) + |\nabla \chi(\x)|^2 g(\x-\y)^2\, |\chi(\y)|^2 \right)d\x \,d\y.
\end{equation*}
Using the eigenvalue equation for $g$ one finds
\begin{equation*}
\tr(-\nabla^2_\x)\gamma = 2 X(\gamma^{1/2}) - \frac14\tr\gamma + \iint |\nabla
\chi(\x)|^2 g(\x-\y)^2\, |\chi(\y)|^2 \,d\x\,d\y\,.
\end{equation*}
The upper bound will follow from this  if we can find functions $\chi_L$ 
(where $L$ is some free parameter) such
that for $\gamma_L$ defined via $\chi_L$,
\begin{equation}\label{eq:choicechi1}
\gamma_L\leq 1, \ {\rm as\ an\ operator},  \qquad{\rm } 
\qquad \tr\gamma_L\to N, 
\end{equation}
\begin{equation}\label{eq:choicechi2}
\iint |\nabla \chi_L(\x)|^2 g(\x-\y)^2\, |\chi_L(\y)|^2 \,d\x \, d\y \to 0, 
\qquad 
{\rm and} \qquad
\qquad D(\rho_{\gamma_L},\rho_{\gamma_L}) \to 0
\end{equation}
as $L\to \infty$. 
We shall choose $\chi_L$ of the form $\chi_L(\x) =
L^{-3/4}\chi(\x/L)$ for a fixed smooth function $\chi\geq 0$
satisfying $\|\chi\|_4^4=N$.

We note that for any $L^2 $ function $\psi$ (and with
$\widehat{\cdots}$ denoting the Fourier transform)
\begin{equation*}
(\psi,\gamma^{1/2}_L\psi)
= (2\pi)^{3/2} \int \widehat g(\p) |(\widehat{ \chi_L\psi)}(\p)|^2\,d\p
\leq (2\pi)^{3/2} \|\widehat g\|_\infty \|\chi_L\|_\infty^2 \|\psi\|^2_2,
\end{equation*}
which is less than or equal to $\|\psi\|^2_2$ for $L$ large, since
$\|\chi_L\|_\infty\to 0$. This implies the first condition in
\eqref{eq:choicechi1}. To check the second one, we write
\begin{equation*}
\tr\gamma_L= (2\pi)^{3/2} \int \widehat{(g^2)}(\p) |
\widehat{(\chi_L^2)}(\p)|^2\,  d \p.
\end{equation*}
Now
$|\widehat{(\chi_L^2)}(\p)|^2=L^3|\widehat{(\chi^2)}(L\p)|^2$,
which converges to $N\delta(\p)$ as
$L\to\infty$ (recall that $\|\chi\|_4^4=N$). Therefore
\begin{equation*}
\tr\gamma_L \to (2\pi)^{3/2} \widehat{(g^2)}(0) N = N.
\end{equation*}

To check conditions \eqref{eq:choicechi2} we estimate (again using
that $\|g\|_2  =1$),
\begin{equation*}
\iint |\nabla \chi_L(\x)|^2 g(\x-\y)^2\, \chi_L(\y)^2\,d\x \,d\y
\leq \|\chi_L\|_\infty^2 \int |\nabla \chi_L(\x)|^2 \,d\x
= L^{-2} \|\chi\|_\infty^2 \|\nabla\chi\|^2.
\end{equation*}
Moreover, 
\begin{equation*}
D(\rho_{\gamma_L},\rho_{\gamma_L}) = \frac1{2L} \iint
\frac{\chi^2(\x)\phi_L(\x)\phi_L(\y)\chi^2(\y)}{|\x-\y|}\,d\x\,d\y
\end{equation*}
where $\phi_L(\x) = L^3 \int
g^2(L(\x-\y))\chi^2(\y)\,d\y$. Since 
$\phi_L(\x)\to\chi^2(\x)$ as $L\to \infty$, we conclude that 
$D(\rho_{\gamma_L},\rho_{\gamma_L})= L^{-1} D[\chi^4] +
o(L^{-1})$ by dominated convergence.

Hence \eqref{eq:choicechi2} holds, and the proof is complete.
\end{proof}

\textit{Remark:} One might ask whether $X(\gamma^{1/2})$ can be
bounded from above in terms of the usual Dirac type estimate for the
exchange energy, $\int\rho_\gamma(\x)^{4/3}\, d\x$ (cf.
\cite{Lieb1981}).  However, this is not the case, as the following
example shows: define $\gamma_L$, as in the proof Proposition
\ref{z0}, by
$\gamma_L^{1/2}(\X,\X')=L^{-3/2}\chi(\x/L)g(\x-\y)\chi(\y/L)q^{-1/2}\delta_{\sigma,\sigma'}$,
and carry out calculations similar to those done above. We find that
\begin{align*}
       X(\gamma_L^{1/2}) & \to \|\chi\|_4^4 \int\frac{|g(\x)|^2}{2|\x|}\,d\x \ , \\
       \int \rho_{\gamma_L}(\x)^{4/3}\,d\x & \sim L^{-1}
       \|\chi\|_{16/3}^{16/3} \
       , \\
       \int \rho_{\gamma_L}(\x)\,dx & \to \|\chi\|_4^4 \ .
\end{align*}
Hence a bound in terms of the $4/3$-norm can not hold. This example
can be traced back to Cioslowski and Pernal
\cite{CioslowskiPernal1999}.


\section{Minimizer in the Case $Z>0$}

We return here, and in the remainder of this paper, to the general
case in which all $Z_j >0$.  We investigate the functional $\Eh$ in
\eqref{ehatfunctional} and the corresponding relaxed minimization
problem given in \eqref{relaxed}.  Our goal is to show that there is
an energy minimizing $\gamma $ for this problem and that its trace is
$\tr \gamma = N$ whenever $N \leq Z= \sum_j Z_j$. The main result of this section
is contained in the following two theorems, whose elaborate proof will
be given in several parts.

\begin{theorem}\label{exmini}
For any $Z>0$ and $N>0$ one has $\widehat E^{\rm M}_\leq (N)<0$ and the infimum
\eqref{relaxed} is attained. 
\end{theorem}

As explained in the introduction, we do not know how to prove that the
minimizer is unique. The strict convexity of the direct energy
$D(\rho_\gamma,\rho_\gamma) $, however, does imply that all minimizing
$\gamma$'s have the same (spin summed) density $\rho_\gamma (\x)$.

\begin{theorem}\label{fulltrace}
Assume that $N\leq Z$. Then a minimizer of \eqref{relaxed} has trace $N$.
\end{theorem}

In particular, this result implies that in the original problem
\eqref{sqrtenergy} the infimum is achieved in case $N\leq Z$. The critical
number $N_c$ mentioned in the introduction is thus at least $Z$.

\subsection{Proof of Theorem~\ref{exmini}}

By Proposition \ref{z0}, the functional $\Eh(\gamma)$ is non-negative,
if $Z=0$. By using a trial density matrix, we will first show that it
assumes negative values as soon as $Z$ is positive.

\begin{lemma}\label{negative}
              For any $Z>0$ and $N>0$ one has $\widehat E^{\rm M}_\leq (N)<0$.
\end{lemma}

\begin{proof}
Without loss of generality we may assume that there is only one nucleus
of charge $Z$ located at the origin $\x=0$.
        We use the same family $\gamma_L$ of trial density matrices as in
        the proof of the upper bound in Proposition \ref{z0}. Using the
        same estimates, we have
\begin{equation}\label{eq:z0trial}
     \mathcal {\widehat E}^{\rm M}(\gamma_L) = -Z \tr |\x|^{-1} \gamma_L
     + \frac 1L  D[\chi^4] +
o(L^{-1})
\qquad \mbox{as $L\to\infty$.}
\end{equation}
Since $L^3 \int g^2(L(\x-\y))\chi^2(\y)\,d\y \to\chi^2(\x)$, we have
$\tr|\x|^{-1}\gamma_L = L^{-1} \int |\x|^{-1} \chi^4(\x) \,d\x +
o(L^{-1})$.  Hence,
\begin{equation}\label{eq:ztrial}
                      \Eh (\gamma_L)
                      =  L^{-1}\left(- Z \int |\x|^{-1} \chi^4(\x) \,d\x + D[\chi^4]\right) +
o(L^{-1})
                      \qquad \mbox{as $L\to\infty$.}
\end{equation}
For $Z>0$ and $N=\|\chi\|_4^4$ small enough, the first term in brackets can
clearly be made negative by an appropriate choice of $\chi$. This
shows that $\eh_\leq (N)<0$ for small $N$, and hence for all $N$.
\end{proof}


\begin{proposition}\label{exmin}
              Let $Z>0$ and $N>0$. There exists a minimizing sequence $\gamma_j$
for \eqref{relaxed} which converges in $\mathfrak S^1$, the space of 
trace-class operators, i.e., there is a $\gamma$ such that $\tr |\gamma_j -\gamma|
\to 0$. 
\end{proposition}

Before giving the proof of this proposition, we collect some useful
auxiliary material.

\begin{lemma}\label{xhardy}
For every $\epsilon>0$ 
\begin{equation}\label{eq:xhardy}
              \iint_{\{|\x-\y|<\epsilon\}}
          \frac{|\gamma^{1/2}(\X,\X')|^2}{|\x-\y|}\,d\X\,d\X'
          \leq 4\epsilon \tr(-\nabla^2)\gamma
\end{equation}
and
\begin{equation}\label{eq:xhardytotal}
              X(\gamma^{1/2})
              \leq \frac\epsilon4 \tr(-\nabla^2)\gamma + \frac1{4\epsilon} 
\tr\gamma\ .
\end{equation}
\end{lemma}

\begin{proof}
The first inequality can be easily deduced from Hardy's
inequality, which states that 
\begin{equation}\label{hard}
-\nabla^2 \geq \frac 1{4|\x|^2}\,.
\end{equation}

     For the second inequality, we use the well known expression for the
ground state energy of the hydrogen atom, namely,
\begin{equation}\label{hydr}
-\nabla^2
      - \frac{z}{|\x|} \geq -\frac{z^2}4 \ ,
\end{equation}
from which it follows (with $z= 2/\epsilon$) that for every $\X'$
\begin{equation}
    \frac 12  \int \frac{|\gamma^{1/2}(\X,\X')|^2}{|\x-\y|}\, d\X \leq 
\frac{\epsilon}{4} \int |\nabla \gamma^{1/2}(\X,\X') |^2 \, d\X 
+\frac{1}{4 \epsilon}   \int |\gamma^{1/2}(\X,\X') |^2\, d\X \ .
\end{equation}
The lemma follows by integrating over $\X'$. 
\end{proof}

\begin{lemma}\label{xlocal}
        Let $\chi(\x)$ satisfy  $|\chi(\x)|\leq 1$. Then
        \begin{equation*}
          X(\chi^*\gamma^{1/2} \chi)
          \leq X((\chi^*\gamma\chi)^{1/2}).
        \end{equation*}
\end{lemma}

\begin{proof} For convenience we introduce the characteristic function
          of a ball of radius $r$ centered at $\z$
\begin{equation}
B_{\z,r}(\x)=
\begin{cases}
       1& |\x-\z|<r\\
       0& |\x-\z|\geq r.
\end{cases}
\end{equation}
Writing the Coulomb kernel  as
\begin{equation}
      \label{eq:fefferman}
       |\x-\y|^{-1} = \frac1\pi \int_0^\infty \int_{\R^3} B_{\z,r}(\x)
       B_{\z,r}(\y) \,d\z\,\frac{dr}{r^5}
\end{equation}
(Fefferman and de la Llave \cite{FeffermandelaLlave1986}), we get
\begin{equation}\label{eq:xdecomposition}
       X(\delta) = \frac1{2\pi} \int_0^\infty \int_{\R^3}
       \tr \left(\delta B_{\z,r} \delta
       B_{\z,r}\right) \,d\z\,\frac{dr}{r^5}.
\end{equation}
It follows from $|\chi|\leq 1$ and the monotonicity of the operator
square root that 
$$\chi^*\gamma^{1/2}\chi=
\left((\chi^*\gamma^{1/2}\chi)( \chi^*\gamma^{1/2}\chi
)\right)^{1/2} \leq (\chi^*\gamma^{1/2} \gamma^{1/2}
\chi)^{1/2} =(\chi^* \gamma\chi)^{1/2}.$$
Hence
\begin{equation*}
       \tr \left(\chi^*\gamma^{1/2}\chi B_{\z,r}
       \chi^*\gamma^{1/2}\chi B_{\z,r}\right) \leq \tr
       \left((\chi^*\gamma\chi)^{1/2} B_{\z,r}
       (\chi^*\gamma^{1/2}\chi)^{1/2} B_{\z,r}\right).
        \end{equation*}
        The assertion follows now from \eqref{eq:xdecomposition}.
\end{proof}

\begin{proof}[Proof of Proposition \ref{exmin}]
        We choose an arbitrary minimizing sequence $\gamma_j$ for
\eqref{relaxed} and, after passing to a subsequence (if necessary), assume
that $\tr\gamma_j\to\tilde N\in[0,N]$. It follows from
\eqref{eq:xhardytotal} and the hydrogen bound, $\tr Z_k|\x - \RR_k|^{-1}\gamma \leq
(Z_k\epsilon/4Z)\tr(-\nabla^2)\gamma +(Z_k Z/\epsilon)\tr\gamma$ that
        \begin{equation}\label{eq:kinbound}
          \frac 12(1-\epsilon)\tr(-\nabla^2)\gamma_j
          \leq \Eh(\gamma_j) + \frac 1\epsilon(Z^2
          + 1/4)\tr\gamma_j.
        \end{equation}
        Hence the sequence $(-\nabla^2+1)^{1/2}\ \gamma_j \
        (-\nabla^2+1)^{1/2}$ is bounded in $\mathfrak S^1$ and, by the
        Banach-Alaoglu theorem (see \cite{LiebLoss2001}) there exists a
        $\gamma$ such that, after passing to a subsequence (if
        necessary), $\tr K\gamma_j\to \tr K\gamma$ for any operator
        $K$ such that $(-\nabla^2+1)^{-1/2}K(-\nabla^2+1)^{-1/2} $ is
        compact.  This compactness condition is satisfied if $K$ is
        simply multiplication by some function $f\in L^p(\R^3)$ for some
        $3/2\leq p<\infty$ (see \cite[section 13.4]{ReedSimon1978}).  In
        this case we have that
\begin{equation}\label{eq:densityconv}
          \int f(\x) \rho_{\gamma_j}(\x) \,d\x
          = \tr f \gamma_j
          \to \tr f \gamma
          = \int f (\x) \rho_\gamma(\x) \,d\x \ .
\end{equation}
In particular, we can take $f$ in \eqref{eq:densityconv} to be the Coulomb 
potential since this potential can be written as the sum of two functions, one
of which is in $ L^p(\R^3)$ and the other in $ L^q(\R^3)$  with
$3/2 < p < 3$ and $3< q < \infty$.

Note that $0\leq\gamma\leq 1$ and, by the lower semicontinuity of the
$\mathfrak S^1$-norm,
\begin{equation*}
          M = \tr\gamma
          \leq \liminf_{j\to\infty}\tr\gamma_j=\tilde N\leq N.
\end{equation*}
We claim that $\gamma\not\equiv 0$ (and hence $M>0$). Indeed, by
Proposition \ref{negative} one has $\mathcal{\widehat
      E}^{\rm M}(\gamma_j)\leq-\epsilon$ for some $\epsilon>0$ and all
sufficiently large $j$. Hence $\tr V_c\gamma_j\geq\epsilon$ and
by \eqref{eq:densityconv} also $\tr V_c\gamma\geq\epsilon$.

Clearly, $\gamma_j\rightharpoonup\gamma$ in the sense of weak operator
convergence. If $M=\tilde N$, then also $\tr\gamma_j\to\tr\gamma$, and
thus $\gamma_j\to\gamma$ in $\mathfrak S^1$ (see Theorem A.6 in
\cite{Simon1979T}) and we are done.

We are thus left with the case $M<\tilde N$. Our strategy will be to
construct a minimizing sequence $\gamma_j^0$ out of the $\gamma_j$
which converges to $\gamma$ in $\mathfrak S^1$. We choose a quadratic
partition of unity, $(\chi^0)^2 + (\chi^1)^2 \equiv 1$, where $\chi^0$
is a smooth, symmetric decreasing function with $\chi^0({\bf 0})=1$,
$\chi^0(\x)<1$ if $|\x|>0$ and $\chi^0(\x)=0$ if $|\x|\geq 2$. For
fixed $j$, $\tr(\chi^0(\x/R))^2\gamma_j$ is a continuous function of
$R$ which increases from $0$ to $\tr\gamma_j$. If we restrict
ourselves to large $j$, then $\tr\gamma_j>M$ and we can choose an
$R_j$ such that $\tr(\chi^0(\x/R_j))^2\gamma_j= M$. We write
$\chi_j^\nu(\x)=\chi^\nu(\x/R_j)$ and
$\gamma_j^\nu=\chi^\nu_j\gamma_j\chi^\nu_j$ for $\nu=0,1$.

        We claim $R_j\to\infty$. To see this,
assume the contrary, namely that there is a subsequence that converges
to some $R<\infty$. Then, for this subsequence, 
$\chi_j^0(\x)^2\to \chi^0(\x/R)^2$ strongly in any $L^p$. Since
$\rho_{\gamma_j} \rightharpoonup \rho_\gamma$ weakly in $L^p$ for $1<p<3$ by
\eqref{eq:densityconv}, one has
        \begin{equation*}
              \int\chi_j^0(\x)^2 \rho_{\gamma_j}(\x) \,d\x \to \int \chi^0(\x/R)^2
\rho_{\gamma}(\x)\,d\x\,.
        \end{equation*}
        But, by definition, the left side is independent of $j$ and equals $\int
\chi_j^0(\x)^2 \rho_{\gamma_j}(\x)\,d\x = M = \int\rho_{\gamma}(\x)\,d\x$. This is a
contradiction, since $\chi^0(\x)^2<1$ almost everywhere and
$\gamma\not\equiv 0$.

Therefore $\lim_{j\to \infty} R_j=\infty$. We note that
$\gamma_j^0\rightharpoonup\gamma$ in the sense of weak operator convergence. (It
suffices to check the weak convergence on functions of compact
support, since the $\gamma_j^0$ remain uniformly bounded.) By
construction, $\tr\gamma_j^0=\tr\gamma$, so that $\gamma_j^0\to\gamma$
in $\mathfrak S^1$ (again by Theorem A.6 in \cite{Simon1979T}) and it
remains to prove that $\gamma_j^0$ is a minimizing sequence.

              For the kinetic energy we use the IMS formula \cite{Cyconetal1987}
              \begin{equation*}
                      \tr(-\nabla^2\gamma_j) = \tr(-\nabla^2\gamma_j^0) +
                      \tr(-\nabla^2\gamma_j^1) -
                      \tr[(|\nabla\chi_j^0|^2+|\nabla\chi_j^1|^2)\gamma_j].
              \end{equation*}
              Since $R_j\to\infty$, one has
$\||\nabla\chi_j^0|^2+|\nabla\chi_j^1|^2\|_\infty\to 0$ and therefore
              \begin{equation}\label{eq:minkin}
                      \tr(-\nabla^2)\gamma_j = \tr(-\nabla^2)\gamma_j^0 + \tr(-\nabla^2)\gamma_j^1 +
o(1).
              \end{equation}
              For the attraction term we use again that $R_j\to\infty$, so
$\tr|\x-{\bf R}_k|^{-1}\gamma_j^1\to 0$ and
              \begin{equation}\label{eq:minattr}
                      \tr|\x-{\bf R}_k|^{-1}\gamma_j = \tr|\x-{\bf R}_k|^{-1}\gamma_j^0 + o(1).
              \end{equation}
              For the repulsion term we use that $\rho_{\gamma_j^0}\leq
\rho_{\gamma_j}$ pointwise and get
              \begin{equation}\label{eq:minrep}
                      D(\rho_{\gamma_j},\rho_{\gamma_j})\geq
D(\rho_{\gamma_j^0},\rho_{\gamma_j^0}).
              \end{equation}

Finally, we turn to the exchange term, which we write as
        \begin{align*}
          X(\gamma_j^{1/2})
          = X(\chi_j^0 \gamma_j^{1/2} \chi_j^0) +
          X(\chi_j^1 \gamma_j^{1/2} \chi_j^1)
          + 2 X(\chi_j^0 \gamma_j^{1/2} \chi_j^1).
        \end{align*}
        We shall show that
        \begin{equation}\label{eq:minx}
                      X(\gamma_j^{1/2})
          \leq X((\gamma_j^0)^{1/2}) + X((\gamma_j^1)^{1/2}) + o(1).
        \end{equation}
        It follows from Lemma \ref{xlocal} that $X(\chi_j^\nu \gamma_j^{1/2}
        \chi_j^\nu) \leq X((\gamma_j^\nu)^{1/2})$. To show that the
        off-diagonal term tends to zero we decompose, for any $\epsilon>0$,
              \begin{align*}
                X(\chi_j^0 \gamma_j^{1/2} \chi_j^1) = &
                \iint_{\{|\x-\y|<\epsilon/2\}}
                \frac{|\chi_j^0(\x)\gamma_j^{1/2}(\X,\X')\chi_j^1(\y)|^2}{2 |\x-\y|}\,d\X\,d\X' \\
                & + \iint_{\{|\x-\y|\geq\epsilon/2\}}
                \frac{|\chi_j^0(\x)\gamma_j^{1/2}(\X,\X')\chi_j^1(\y)|^2}{2
                  |\x-\y|}\,d\X\,d\X'.
        \end{align*}
        The term with the singularity is controlled by \eqref{eq:xhardy},
        \begin{align*}
              \iint_{\{|\x-\y|<\epsilon/2\}}
          \frac{|\chi_j^0(\x)\gamma_j^{1/2}(\X,\X')\chi_j^1(\y)|^2}{2 |\x-\y|}\,d\X\,d\X'
          & \leq \epsilon
\tr(-\nabla^2)\chi_j^0\gamma_j^{1/2}(\chi_j^1)^2\gamma_j^{1/2}\chi_j^0
\\
          & \leq \epsilon \tr(-\nabla^2)\chi_j^0\gamma_j\chi_j^0 .
              \end{align*}
\renewcommand{\thefootnote}{${1}$}
              This can be made arbitrarily small be choosing $\epsilon$
              small. We pick some $\delta>0$ and 
              decompose the term without singularity into two pieces, depending on whether 
$|\y|<\delta R_j$ or not. In the first case we estimate$^1$ \footnotetext{The following two paragraphs slightly differ from the published version in Phys. Rev. A \textbf{76} (2007), 052517. We are grateful to M. Tiefenbeck for pointing out an error at this point of the proof.}
\begin{align}\label{neweq}
& \iint_{\{|\x-\y|\geq \epsilon/2, \, |\y|<\delta R_j \}} \frac{|\chi_j^0(\x)\gamma_j^{1/2}(\X,\X')\chi_j^1(\y)|^2}{2|\x-\y|} \,d\X d\X' \\ \nonumber
& \qquad \leq \epsilon^{-1} \iint_{\{|\y|<\delta R_j \}} |\gamma_j^{1/2}(\X,\X')\chi_j^1(\y)|^2 \,d\X d\X' \\ \nonumber 
& \qquad = \epsilon^{-1} \tr\chi_{\{|\x|<\delta R_j\}} (\chi_j^1)^2 \gamma_j \\ \nonumber
& \qquad \leq \epsilon^{-1} N \|\chi_{\{|\x|<\delta R_j\}} \chi_j^1\|_\infty^2 \,.
\end{align}
Since $\chi^1$ is smooth with $\chi^1(0)=0$, the supremum-norm of the function $\chi_{\{|\x|<\delta R_j\}} \chi_j^1$ (which is independent of $R_j$ by scaling) can be made arbitrarily small by choosing $\delta$ small. Hence the double integral \eqref{neweq} can be made arbitrarily small. 

In the complementary region one may argue as follows. We pick some $A$ and choose $j$ so large that $R_j> \delta^{-1} A$. By estimating $|\x-\y|\geq \delta R_j -A$ if $|\x|<A$ and $|\y|>\delta R_j$, we obtain
              \begin{align*}
                      & \iint_{\{|\x-\y|\geq\epsilon/2,\, |\y|\geq \delta R_j \}}
          \frac{|\chi_j^0(\x)\gamma_j^{1/2}(\X,\X')\chi_j^1(\y)|^2}{2 |\x-\y|}\,d\X\,d\X' \\
                      & \qquad \leq \iint_{\{|\x-\y|\geq\epsilon/2, |\x|\geq A \}}
          \frac{|\chi_j^0(\x)\gamma_j^{1/2}(\X,\X')|^2}{2 |\x-\y|}\,d\X\,d\X'
                      + \iint_{\{|\x|< A,\, |\y|\geq \delta R_j \}}
          \frac{|\gamma_j^{1/2}(\X,\X')|^2}{2 |\x-\y|}\,d\X\,d\X' \\
          & \qquad \leq \epsilon^{-1} \iint_{\{|\x|\geq A \}}
          \chi_j^0(\x)^2|\gamma_j^{1/2}(\X,\X')|^2 \,d\X\,d\X'
                      + (2(\delta R_j-A))^{-1} \iint |\gamma_j^{1/2}(\X,\X')|^2 \,d\X\,d\X' \\
                      & \qquad = \epsilon^{-1} \tr \chi_{\{|\x|\geq A \}}\gamma_j^0
          + (2(\delta R_j-A))^{-1} \tr\gamma_j.
        \end{align*}
        Since $\gamma_j^0\to\gamma$ in $\mathfrak S^1$, one
has $\tr \chi_{\{|\x|\geq A \}}\gamma_j^0 \to \tr\chi_{\{|\x|\geq A
\}}\gamma$. This can be made arbitrarily small by choosing $A$ large.
Since $R_j\to\infty$, the term $(2(\delta R_j-A))^{-1} \tr\gamma_j$ converges
to $0$. This proves \eqref{eq:minx}.

              Collecting \eqref{eq:minkin}--\eqref{eq:minx} we find that
              \begin{equation*}
                      \mathcal{\widehat E}^{\rm M}(\gamma_j)
                      \geq \mathcal{\widehat E}^{\rm M}(\gamma_j^0) +
                      \left( -\frac 12 \tr \nabla^2 \gamma_j^1 - X(\gamma_j^1)+ \frac1{8}\tr\gamma_j^1 \right)
                      + o(1).
              \end{equation*}
              We have shown in the proof of Proposition \ref{z0} that
              the term in brackets is non-negative. Hence
              \begin{equation*}
                      \liminf_{j\to\infty} \mathcal{\widehat E}^{\rm M}(\gamma_j)
                      \geq \liminf_{j\to\infty} \mathcal{\widehat E}^{\rm M}(\gamma_j^0),
              \end{equation*}
              which shows that $\gamma_j^0$ is a minimizing sequence. This concludes
the proof.
\end{proof}


\begin{proposition}\label{lsc}
        Let $\gamma_j\to\gamma$ in $\mathfrak S^1$. Then
        \begin{equation}\label{eq:lsc}
              \liminf_{j\to\infty} \mathcal{\widehat E}^{\rm M}(\gamma_j) \geq \mathcal{\widehat
E}^{\rm M}(\gamma).
        \end{equation}
\end{proposition}

\begin{proof}
              The bound \eqref{eq:kinbound} shows that $E=\liminf_{j\to\infty}
\mathcal{\widehat E}^{\rm M}(\gamma_j)>-\infty$. Moreover, we may assume that
$E<\infty$, for otherwise there is nothing to prove. After passing to a
subsequence (if necessary), we may assume that $\mathcal{\widehat
E}^{\rm M}(\gamma_j)\to E$. As in the proof of Proposition \ref{exmin} there
exists a $\gamma$ such that, after passing to a subsequence if necessary,
$\tr K\gamma_j\to \tr K\gamma$ for any operator $K$ such that
$(-\nabla^2+1)^{-1/2}K(-\nabla^2+1)^{-1/2}$ is compact. In
particular, \eqref{eq:densityconv} holds. By weak lower-semicontinuity we
infer that
              \begin{equation}\label{eq:lsckin}
              \tr\left(-\half \nabla^2+1/8\right)\gamma
              \leq \liminf_{j\to\infty} \tr\left(-\half\nabla^2+1/8\right)\gamma_j.
        \end{equation}

        Now we turn to the repulsion term. Since
$D(\rho_{\gamma_j},\rho_{\gamma_j})$ is
        bounded we may, passing to a subsequence (if necessary), assume
        that $\rho_{\gamma_j}$ converges weakly to some $\rho$ with respect
        to the $D$-scalar product. With the help of \eqref{eq:densityconv} one
concludes that $\rho=\rho_\gamma$.
        Weak lower-semicontinuity with respect to the $D$-norm implies that
        \begin{equation}\label{eq:lscrep}
              D(\rho_\gamma,\rho_\gamma)
              \leq \liminf_{j\to\infty} D(\rho_{\gamma_j},\rho_{\gamma_j}).
        \end{equation}
        The continuity of the attraction term follows from
\eqref{eq:densityconv}, since $|\x|^{-1}\in L^{3/2} + L^p$ for $p>3$, therefore
        \begin{equation}\label{eq:lscattr}
              \lim_{j\to\infty}\tr V_c\gamma_j=\tr V_c\gamma \ .
        \end{equation}
        Finally, we prove continuity of the exchange term. Similarly as in the
proof of Proposition \ref{exmin} we decompose, for any $\epsilon>0$,
        \begin{align*}
          |X(\gamma_j^{1/2}) - X(\gamma^{1/2})|
          & \leq \iint_{\{|\x-\y|<\epsilon/2\}}
          \frac{|\gamma_j^{1/2}(\X,\X')|^2 + |\gamma^{1/2}(\X,\X')|^2}{2
|\x-\y|}\,d\X\,d\X' \\
          & \qquad + \iint_{\{|\x-\y|\geq\epsilon/2\}}
          \frac{\left| |\gamma_j^{1/2}(\X,\X')|^2
          -|\gamma^{1/2}(\X,\X')|^2\right|}{2 |\x-\y|}\,d\X\,d\X'
        \end{align*}
        According to Lemma \ref{xhardy} the term involving the singularity is
bounded by $\epsilon \tr(-\nabla^2)(\gamma_j + \gamma )$, which can be
made arbitrarily small (recall that $\tr[-\nabla^2(\gamma_j + \gamma)]$ is
bounded). To treat the term without the singularity we use the fact that
the mapping $K\mapsto |K|^{1/2}$ is continuous from $\mathfrak S^1$ to
$\mathfrak S^2$ (see Example 2 after Theorem 2.21 in \cite{Simon1979T}). Hence
$\gamma_j^{1/2}\to\gamma^{1/2}$ in $\mathfrak S^2$, and we can bound
\begin{multline}\nonumber
                   \left(    \iint_{\{|\x-\y|\geq\epsilon/2\}}
          \frac{||\gamma_j^{1/2} (\X,\X')|^2
          -|\gamma^{1/2}(\X,\X')|^2|}{2 |\x-\y|}\,d\X\,d\X' \right)^2\\
          \leq
          {\iint \left|\gamma_j^{1/2}(\X,\X') - \gamma^{1/2}(\X,\X')\right|^2
\,d\X\,d\X' \iint_{\{|\x-\y|\geq\epsilon/2\}}
          \frac{( |\gamma_j^{1/2}(\X,\X')| + |\gamma^{1/2}(\X,\X')|)^2 }{4
|\x-\y|^2}\,d\X'\,d\X'} \\
         \leq
          \|\gamma_j^{1/2} - \gamma^{1/2}\|_2^2\
          2 \epsilon^{-2} \tr(\gamma_j+\gamma) .
        \end{multline}
        The first factor tends to zero by the   convergence of
$\gamma_j^{1/2}$ mentioned before, 
and the second one remains bounded. Hence we have
proved that
              \begin{equation}\label{eq:lscx}
                      \lim_{j\to\infty} X(\gamma_j^{1/2}) = X(\gamma^{1/2}).
              \end{equation}
By collecting \eqref{eq:lsckin}--\eqref{eq:lscx} we arrive at \eqref{eq:lsc}.
\end{proof}

\begin{proof}[Proof of Theorem~\ref{exmini}]
According to Proposition~\ref{exmin}, there exists a minimizing sequence
that converges strongly to some $\gamma$. By Proposition~\ref{lsc}, this
$\gamma$ is a minimizer of $\Eh$.
\end{proof}

\subsection{Proof of Theorem~\ref{fulltrace}}

Assume that $N\leq Z$. Under this assumption we shall show that a $\g$
minimizing $\Eh(\gamma)$ satisfies $\tr\gamma =N$.

Assuming the contrary, we shall find a trace class operator $\sigma\geq
0$ such that for
$\gamma_\epsilon=(1-\epsilon\|\sigma\|)\gamma+\epsilon\sigma$ and all
sufficiently small $\epsilon>0$,
              \begin{equation}\label{eq:fulltrace}
                      \mathcal{\widehat E}^{\rm M}(\gamma_\epsilon)<\mathcal{\widehat E}^{\rm M}(\gamma) \ .
              \end{equation}
The factor  $(1-\epsilon\|\sigma\|)$ guarantees that 
$0\leq \gamma_\epsilon\leq 1$ for
$0<\epsilon\leq\|\sigma\|^{-1}$.  If $\tr \gamma< N$, which we  assume,
then also $\tr \gamma_\epsilon <N$ for small $\epsilon$ and \eqref{eq:fulltrace}
leads to a contradiction since $\gamma$ was assumed to be a minimizer.

              To prove \eqref{eq:fulltrace} we use convexity for the homogeneous terms
in the functional $\mathcal{\widehat E}^{\rm M}$ and expand the repulsion term
explicitly. This leads to
              \begin{equation}\label{eq:fulltrace1}
                      \mathcal{\widehat E}^{\rm M}(\gamma_\epsilon)
                      \leq \mathcal{\widehat E}^{\rm M}(\gamma) +
\epsilon\left(\tr(-\nabla^2-\phi_\gamma+1/8)\sigma -
X(\sigma^{1/2})\right)
                      - \epsilon R_1 + \epsilon^2 R_2 \ ,
\end{equation}
where
              \begin{align*}
                \phi_\gamma(\x)& = V_c(\x)-\int\frac{\rho_\gamma(\y)}{|\x-\y|}\,d\y \
                , \\
                R_1 & = \|\sigma\| \left(\mathcal{\widehat E}^{\rm M}(\gamma) +
                  D(\rho_\gamma,\rho_\gamma) \right) \ , \\
                R_2 & =
D(\rho_\sigma-\|\sigma\|\rho_\gamma,\rho_\sigma-\|\sigma\|
\rho_\gamma)
                \ .
              \end{align*}
              Now we proceed similarly as in the proof of Proposition \ref{z0}, letting
$\sigma=\sigma_L$ depend on a (large) parameter $L$. More precisely, we
define $\sigma_L$ by
              \begin{equation}\label{eq:sigmal}
                      \sigma_L^{1/2}(\X,\X') = L^{-3/2}\chi(\x/L)
                      g(\x-\y) \chi(\y/L)\, q^{-1/2} \delta_{\sigma,\sigma'} \ ,
              \end{equation}
              with $g$ as in \eqref{eq:g} and $\chi\geq 0$ a smooth function satisfying
$\|\chi\|_4^4=1$. Asymptotically, for large $|\x|$,
$\phi_\gamma(\x) \approx  (Z-   \tr \gamma )|\x|^{-1}$, which is positive
by our assumption. It follows similarly to the proof of Proposition
\ref{negative} that
              \begin{equation*}
                      \tr(-\nabla^2-\phi_\gamma+1/8)\sigma_L - X(\sigma_L^{1/2})
             =  - \frac{Z-\tr \gamma }L \int |\x|^{-1} 
\chi^4(\x) \,d\x + o(L^{-1}) \ .
              \end{equation*}

It remains to show that the terms $R_1$ and $R_2$ are relatively small.
In the proof of Proposition \ref{z0} and in \eqref{eq:z0trial} we
showed that $\|\sigma_L\|=\mathcal O(L^{-3})$ and
$D(\rho_{\sigma_L},\rho_{\sigma_L})=\mathcal O(L^{-1})$, which implies
that $R_1= \mathcal O(L^{-3})$ and $R_2= \mathcal O(L^{-1})$. We can then
choose $L$ large enough and $\epsilon$ small enough to conclude
\eqref{eq:fulltrace}.

This finishes the proof of Theorem~\ref{fulltrace}.

\section{Further Properties}

\subsection{Properties of the Minimal Energy}

Recall that $E^{\rm M}(N)$ as defined in (\ref{sqrtenergy}) is the lowest energy
of $\E^{\rm M} (\gamma) $ under the condition $\tr\gamma=N$. 
This energy is closely
related to $\eh_\leq(N)$ defined in (\ref{relaxed}).

\begin{proposition}\label{minz=}
For any $Z>0$ and $N>0$ one has $E^{\rm M}(N) = \widehat E^{\rm M}_\leq (N)-N/8$.
\end{proposition}

What this proposition really says is that $E^{\rm M}(N) + N/8$ is a
monotone non-decreasing function of $N$. This, in turn, follows from
the fact that we can always add mass
$\delta N$ far away from the nuclei, with an energy as close as we
please to $-\delta N/8$. This was shown  in the proof of Theorem \ref{fulltrace},
and we shall not repeat the argument.


\begin{proposition}
              For any $Z>0$ the energies $\widehat E^{\rm M}_\leq(N)$ and $E^{\rm M}(N)$ are convex
functions of $N$. They are strictly convex for $0<N\leq Z$.
\end{proposition}

\begin{proof}
              By Proposition \ref{minz=} it suffices to consider $\widehat E^{\rm M}_\leq(N)$.
The convexity follows from the convexity of the functional. Moreover,
from Theorem \ref{fulltrace} we know that minimizers for $0<N<N'\leq Z$ have
different traces, and hence different densities. The {\emph strict}
convexity follows hence from the strict convexity of $D(\rho,\rho)$ in
$\rho$.
\end{proof}

We now  prove that the energy is bounded from below uniformly in $N$ for
fixed $Z$.

\begin{proposition}
              There is  a constant $C>0$ (independent of $N$ and the charges and positions of the nuclei) such that for all $Z>0$ and $N>0$, $\widehat E^{\rm M}_\leq
(N)\geq -CZ^3$.
\end{proposition}

{\it Remark:}
The proof below does {\it not} use the property that $\gamma\leq
1$ and this results in the exponent $3$ , which is not optimal in the
fermionic case.  Without the restriction $\gamma \leq
1$, the exponent 3 is optimal, however.

\begin{proof} First, let us consider the atomic case with a nucleus
of charge $Z$ located at the origin ${\bf R} =0$.
              We consider $\psi(\X,\X')=\gamma^{1/2}(\X,\X')$ as a wave function in
$L^2(\R^6)$ and find after symmetrization
              \begin{equation*}
                      \mathcal{\widehat E}^{\rm M}(\gamma) = \frac12
                      \left\bra\psi \left|
-\frac 12\nabla^2_\x-\frac12\nabla^2_\y-Z|\x|^{-1}
-Z|\y|^{-1}-\frac 1{|\x-\y|}+\frac 14\right| \psi\right\ket +
D(\rho_\gamma,\rho_\gamma).
              \end{equation*}
              By the positive definiteness of the Coulomb kernel,
$D(\rho_\gamma,\rho_\gamma)\geq 2D(\rho_\gamma,\sigma_Z) -
D(\sigma_Z,\sigma_Z)$ for any $\sigma_Z$. Hence
              \begin{equation*}
                      \mathcal{\widehat E}^{\rm M}(\gamma) \geq \frac12
                      \left\bra\psi\left|- \frac12
\nabla^2_\x-\frac 12\nabla^2_\y-V_{Z}(\x)-V_{Z}(\y)-\frac
1{|\x-\y|}+\frac 14\right|\psi\right\ket -
D(\sigma_Z,\sigma_Z)
              \end{equation*}
              with $V_{Z}(\x) = Z|\x|^{-1} -
           \int |\x- \x'|^{-1} \sigma_Z (\x') d\x'  $.
We shall choose
$\sigma_Z$ in such a way that
              \begin{equation}\label{eq:nobs}
                      -\frac
                      12\nabla^2_\x-\frac12\nabla^2_\y-V_{Z}(\x)-V_{Z}(\y)-\frac 1{|\x-\y|}+\frac 14\geq 0.
              \end{equation}
              From this it follows that $\mathcal{\widehat E}^{\rm M}(\gamma) \geq
-D(\sigma_Z,\sigma_Z)$. Actually, we shall choose $\sigma_Z$ of the form
$\sigma_Z(\x)=Z^4\sigma(Z\x)$ for some fixed $\sigma$, which yields $D(\sigma_Z,\sigma_Z)= Z^3
D(\sigma,\sigma)$.

              To prove \eqref{eq:nobs} we make an orthogonal change of variables,
$\s=(\x-\y)/\sqrt2$, $\tb=(\x+\y)/\sqrt2$, so that the operator on the left
side of \eqref{eq:nobs} becomes
              \begin{equation*}
                      \left(-\frac12\nabla^2_\s- \frac 1{\sqrt2 |\s|}+\frac
                        14\right) + \frac14\left(-\nabla^2_\tb-
4V_Z((\tb+\s)/\sqrt2)\right) + \frac14\left(-\nabla^2_\tb- 4V_Z((\tb-\s)/\sqrt2)\right).
              \end{equation*}
              The operator in the first brackets is non-negative (see
              Eq.~(\ref{hydr})). Hence it suffices to
choose $\sigma$ such that the operator $-\nabla^2_\tb- 4V_Z((\tb+\ab)/\sqrt2)$ is
non-negative of any $\ab\in\R^3$. 
Note that $V_{Z}(\tb) = Z^2 V(Z\tb)$ with $V(\x)
= |\x|^{-1} - \int |\x- \x'|^{-1} \sigma (\x') d\x'$.
After scaling and translation, we
have to prove that $-\nabla^2_\x-8V(\x)\geq 0$. For this we choose $\sigma$ a
non-negative, spherically symmetric function with $\int \sigma\,dx =1$
and with support in $\{|\x|\leq 1/32\}$. Then by Newton's theorem $V(\x)=0$
for $|\x|\geq 1/32 $, and for $|\x|\leq 1/32 $ one has $8V(\x)\leq 1/(4|\x|^2)$,
so $-\nabla^2_\x-8V(\x)\geq 0$ by Hardy's inequality \eqref{hard}. This concludes the
proof in the atomic case.

In the molecular case we proceed as follows: We recall that we are not
taking account of the (fixed) nuclear repulsion $U$, and this means
that we can freely place the nuclei at locations that minimizes the
energy $\eh(N)$.  We assert that the best choice of the ${\bf R}_j$ is
one in which they are all equal and, by translation invariance, this
common point can be the origin. The problem thus reduces to the atomic
case with a nucleus whose charge is the total charge $Z$. That the
optimum choice is equal ${\bf R}_j$ follows from the fact that for
{\it any} $\gamma$ the attractive energy for nucleus $j$ is $-\int
\rho_\gamma (\x) |\x-{\bf R}_j|^{-1} d\x$ and the best possible energy
is obtained by placing all the ${\bf R}_j $ at the point ${\bf R}$
that maximizes this integral.
\end{proof}

\subsection{Properties of the Minimizer}  \label{minprop}

\begin{proposition} \label{nullspace}
              Let $\gamma$ be a minimizer of \eqref{relaxed} and let $M_\gamma=\{\x :
\rho_\gamma(\x)>0 \}$. Then the null-space of the spin-summed density matrix,
$\ker\, \tr_\sigma\gamma$,  coincides with the set of $L^2(\R^3)$ functions 
that vanish identically on $M_\gamma$. 
\end{proposition}

Another way to say this is that if $ \tr_\sigma\gamma$ has a zero eigenvalue then
the eigenfunction vanishes wherever the density $\rho_\gamma$ is non-zero.
     In particular, if $\rho_\gamma>0$ almost everywhere  then $0$ is not an
eigenvalue of the spin-summed density matrix $\tr_\sigma\gamma$.

\begin{proof}
      Write $(\tr_\sigma\gamma)(\x,\x')=\sum_j\lambda_j
      \psi_j(\x)\psi(\x')^*$ with $\psi_j$ orthonormal and
      $0<\lambda_j\leq q$. Then $\R^3\setminus M_\gamma=\bigcap_j \{\x :
       \psi_j(\x)= 0 \}$, and if $\phi=0$ a.e. on
      $M_\gamma$ then obviously $\gamma\phi\equiv 0$. Conversely, let
      $\phi\in\ker\, \tr_\sigma\gamma$ and consider
      \begin{equation*}
        \gamma_\epsilon= \tr_\sigma\gamma + \epsilon \left(|\phi\rangle\langle \phi| -
        |\psi_1\rangle\langle \psi_1|\right).
      \end{equation*}
      One has $\tr\gamma_\epsilon=\tr\gamma\leq N$,
$0\leq\gamma_\epsilon\leq 1$ for $0\leq\epsilon\leq \lambda_1$ and
      \begin{equation*}
        \gamma_\epsilon^{1/2} = \big(\tr_\sigma\gamma\big)^{1/2} + \sqrt\epsilon |\phi\ket\bra\phi|
        + \left(\sqrt{\lambda_1-\epsilon}-\sqrt\lambda\right)
        |\psi_1\ket\bra\psi_1| \ .
      \end{equation*}
As noted in the introduction, it follows from convexity that
minimizing $\mathcal{\widehat E}^{\rm M}$ for density matrices $0\leq
\gamma\leq 1$ with $q$ spin states is equivalent to minimizing under
the condition $0\leq \gamma\leq q$ without spin. Hence
      \begin{equation*}
        E_\leq ^{\rm M}(N) \leq \mathcal{\widehat E}^{\rm M}(\gamma_\epsilon)
        = \mathcal{\widehat E}^{\rm M}(\gamma) - \sqrt\epsilon C[\phi] +\mathcal O(\epsilon),
      \end{equation*}
      where
      \begin{equation*}
        C[\phi] = \iint \frac{\phi(\x)^*\gamma^{1/2}(\x,\x')\phi(\x')}{|\x-\y|}\,d\x\,d\x'
        = \sum_j\sqrt\lambda_j \iint
        \frac{\phi(\x)^*\psi_j(\x)\psi_j(\x')^*\phi(\x')}{|\x-\y|}\,d\x\,d\x'
        \geq 0\ .
      \end{equation*}
              Since $\gamma$ is a minimizer, one has $C[\phi]=0$, which by the positive
definiteness of the Coulomb kernel means $\phi \psi_j^*=0$ a.e. for
all $j$. Hence $\phi=0$ a.e. on $M_\gamma$.
              \end{proof}

At the other end of the spectrum of $\gamma$, we 
comment on the eigenvalue $1$ of the minimizer. Consider the
minimization problem \eqref{relaxed} without the constraint $\gamma\leq
1$,
\begin{equation}\label{eq:minzb}
        \widehat E^{\rm boson}_\leq (N)= \inf\{ \mathcal{\widehat E}^{\rm M}(\gamma) :\
        \gamma\geq 0, \,\tr\gamma\leq N\} \ .
\end{equation}
This energy can be interpreted as the ground state energy of $N$ {\it
      bosons} in the M\"uller model.  Obviously, $\widehat E^{\rm
      boson}_\leq (N)\leq \widehat E^{\rm M}_\leq (N)$ with equality for
$N\leq 1$. For large values of $N$ we expect them to differ, however.

\begin{proposition}\label{propq}
Assume that $\widehat E^{\rm boson}_\leq (N)< \widehat E^{\rm M}_\leq
(N)$ for some $N$ and $Z$. Then any minimizer $\gamma$ of
\eqref{relaxed} has at least one eigenvalue $1$.
\end{proposition}

\begin{proof}
    Assume, on the contrary, that $\gamma<1$ and let $\gamma_b$ denote a
    minimizer for \eqref{eq:minzb}. (The existence is shown in the same
    way as in the proof of Theorem \ref{exmini}.) Then $\gamma_\epsilon
    =(1-\epsilon)\gamma+\epsilon\gamma_b$ satisfies
    $\tr\gamma_\epsilon\leq N$ and $0\leq\gamma_\epsilon\leq 1$ for
    sufficiently small $\epsilon>0$.  Moreover, by convexity,
              \begin{equation*}
                      \mathcal{\widehat E}^{\rm M}(\gamma_\epsilon)
                      \leq (1-\epsilon) \widehat E^{\rm M}_\leq (N) +
                      \epsilon \widehat E^{\rm boson}_\leq (N)
                      < \widehat E^{\rm M}_\leq (N)\ ,
              \end{equation*}
              contradicting the fact that $\gamma$ is a minimizer.
\end{proof}

It is not difficult to see that $\widehat E^{\rm M}_\leq (N) \sim N^{1/3}
Z^2$ for large $N$ and $Z$, while $\widehat E^{\rm boson}_\leq (N)\sim
N Z^2$. Hence clearly $\widehat E^{\rm boson}_\leq (N)< \widehat
E^{\rm M}_\leq (N)$ for large $N$ and $Z$.

Lathiotakis et al. \cite{Lathiotakisetal2007} find numerically that in
fact occupation numbers that correspond to core electrons of large
atoms all have the value one.

\begin{proposition}\label{realmin} Let $\gamma(\X, \X')$
    be a minimizer of $\Eh(\gamma)$ for some $N$ and let
    $\widehat\gamma(\x,\x') = \sum_\sigma \gamma(\x,\sigma, \y,\sigma) $ be the
    spin-summed minimizer. Then $\widehat\gamma(\x,\x') $ is necessarily real.
\end{proposition}

\begin{proof}
    It suffices to show that $\widehat\g^{1/2}$ is real. Write
    $\widehat\g^{1/2}(\x,\y) = A(\x,\x') + i B(\x,\x')$, where $A$ is
    real and symmetric and $B$ is real and antisymmetric, whence $iB$ is
    self adjoint.  Define $\delta =A^2-B^2$, noting that both $A^2 $ and
    $-B^2$ are positive (semidefinite).  The kinetic and potential
    energy of $\delta $ and $\g$ are equal. Moreover, the densities
    $\rho_\gamma(\x)$ and $\rho_\delta(\x)$ are equal. Therefore, we
    just have to show that the exchange terms favor $\delta$, i.e.,
    $X(\delta^{1/2}) > X(\g^{1/2})$.

    To prove this assertion use the concavity of $X(\cdot)$ to conclude
    that $X(\delta^{1/2}) \geq X(|A|) +X(|B|)$, where $|A| = \sqrt{A^2}$
    and $|B| = \sqrt{B^\dagger B } =\sqrt{-B^2}$. On the other hand
    $X(\g^{1/2} )= X(A) +X(B)$, with the obvious meaning that $X(A) =
    \half \int |A(\x,\x')|^2 |\x-\x'|^{-1} d\x\, d\y$ and similarly for
    $X(B)$.

To conclude the proof we have to show that $X(|A|) \geq X(A)$ and
$X(|B|) > X(B)$ if $B\neq 0$. For the first, we write $A=A_+ - A_-$
and $|A| = A_+ +A_-$, where $A_\pm$ are both positive operators.
Clearly, the cross term $\int A_+(\x,\x') A_-(\x', \x) |\x-\x'|^{-1}
d\x\, d\y \geq 0$ since $|\x-\x'|$ is positive definite. The same
argument applies to $iB=B_+ - B_-$, but now we want to show that $\int
   B_+(\x,\x') B_-(\x', \x) |\x-\x'|^{-1} d\x\, d\y >0$ unless $B=0$.

   To show this we use the fact that the positive definiteness of the
   Coulomb kernel implies that $\int \alpha (\x,\x') \beta (\x', \x) )
   |\x-\x'|^{-1} d\x\, d\y$ is (operator) monotone in $\alpha$ and in
   $\beta$. Therefore, it suffices to show positivity for selected
   eigenfunctions of $B_\pm$. That is, we replace $B_+(\x,\x') $ by
   eigenfunctions $\phi_+(\x) \phi_+(\x')^*$ and similarly we replace
   $B_-(\x,\x') $ by $\phi_-(\x) \phi_-(\x')^*$.

   Since $iB$ is imaginary and antisymmetric, however, its positive and
   negative spectra are equal, apart from sign, so $B_\pm$ have the same
   spectrum. Moreover, $B_\pm$ are complex conjugates of each other.
   Therefore, for every $\phi_+(\x)$ there is a $\phi_-(\x)$ and the two
   functions are complex conjugates of each other.  In short, it
   suffices to show strict positivity of $\int \phi(\x)^2
   (\phi(\x')^*)^2 |\x-\x'|^{-1} d\x\, d\y$, but this is true as long as
   the function $\phi $ is not identically zero (since the Coulomb
   kernel is positive definite).
\end{proof}

Finally, we show that a minimizer of $\Eh(\gamma)$ satisfies the variational equation~\eqref{vareq}, as claimed in the Introduction.

\begin{proposition}\label{lagrange2}
Let $\gamma$ be a minimizer of $\Eh(\gamma)$. Then
\begin{equation}\label{vareq2}
\left(-\half \nabla_\x^2 - \half \nabla_\y^2 - \varphi_\gamma(\x) - \varphi_\gamma(\y)
- \frac 1{|\x-\y|} - 2\mu\right) \gamma^{1/2}(\X,\X') = 
\sum_i 2 e_i \psi_i(\X) \psi_i(\X')^*\,.
\end{equation}
Here, $\varphi_\gamma(\x)= V_c(\x) - \int \rho_\gamma(\y) |\x-\y|^{-1}
d\y$ denotes the effective potential, $\mu\leq -1/8$ is the chemical
potential, $e_i\leq 0$ and the $\psi_i(\X)$ are eigenfunctions of
$\gamma$ with eigenvalue $1$.
\end{proposition}

\begin{proof}
    Let $\mu$ be the slope of a tangent to the curve $E^{\rm M}(N)$ at
    $N$.  Since $E^{\rm M}(N)$ is convex, such a tangent always exists,
    although it may not be unique in case the derivative of $E^{\rm
      M}(N)$ is discontinuous at this point.

    Since $\gamma$ is a minimizer of $\Eh(\gamma)$, its square-root
    $\gamma^{1/2}$ minimizes the expression
\begin{equation}\label{dfu}
{\mathcal F}(\delta) = \tr\left( -\half \nabla^2 - V_c(\x) -\mu\right) \delta^2 +
D(\rho_{\delta^2},\rho_{\delta^2}) - X(\delta)
\end{equation}
    among {\it all} $\delta$ with $0\leq \delta \leq 1$, irrespective of the
    trace of $\delta^2$. In fact, it is even a
    minimizer if one relaxes the condition $\delta \geq 0$. This follows
    from the fact that $X(\delta)\leq X(|\delta|)$ for any
    self-adjoint operator $\delta$, which was shown in the proof of
    the previous proposition \ref{realmin}.

    Consequently, $\g^{1/2}$ is a minimizer of \eqref{dfu} subject to the
      constraint $-1\leq \delta\leq 1$. From this we conclude that for any
      self-adjoint $\sigma$ with finite trace such that, for small
      $\epsilon$, $\gamma^{1/2}+\epsilon \sigma \leq 1 +$ terms of order
      $\epsilon^2$,
\begin{equation}
   \left. \frac{d}{d\epsilon}  {\mathcal F}(\gamma^{1/2}+\epsilon \sigma)  \right|_{\epsilon=0} \geq 0\,. 
\end{equation}
The derivative can easily be calculated to be
\begin{equation}
    \tr \left[\left( (-\half \nabla^2 - \varphi_\gamma) \gamma^{1/2} 
+ \gamma^{1/2} (-\half \nabla^2-\varphi_\gamma) - Z_\g - 2 \mu\g^{1/2} \right)\, 
\sigma \right]\,,
\end{equation}
where $Z_\gamma$ is defined in \eqref{zgamma}.  The condition on
$\sigma$ is that $\langle\psi_i|\sigma|\psi_i\rangle\leq 0$ for all
$|\psi_i\rangle$ with $\gamma|\psi_i\rangle = |\psi_i\rangle$.  Hence
we conclude that
\begin{equation}
     (-\half \nabla^2 - \varphi_\gamma) \gamma^{1/2} 
+ \gamma^{1/2} (-\half \nabla^2-\varphi_\gamma) - Z_\g - 2 \mu\g^{1/2} 
= \sum_i 2e_i |\psi_i\rangle\langle\psi_i|\,,
\end{equation}
with $e_i\leq 0$.
\end{proof}

The variational equation~\eqref{vareq2} was obtained by varying
$\gamma^{1/2}$ instead of $\gamma$. If $\g$ does not have a zero
eigenvalue (which, for a  spin-invariant minimizer $\gamma$, is the case if
$\rho_\g$ does not vanish on a set of  positive measure, see
Prop.~\ref{nullspace}), then these variations are equivalent. Hence we
conclude that \eqref{vareq2} is actually {\it equivalent} to $\g$
being a minimizer in case $\g$ has no zero eigenvalue. (See the
discussion in Section~\ref{subsec:Mueller}).


\subsection{Virial Theorem}
A well known property of Coulomb systems is the virial theorem, which
quantifies a relation between the kinetic and potential energies. We state it here 
for an atom.

\begin{proposition}\label{virial}
Let $K=1$ (i.e., consider an atom) and let $\gamma$ be a minimizer for $\widehat E^{\rm M}
_\leq (N)$. 
Then
\begin{equation}\label{eq:virial}
2 \tr(-\half\nabla^2\gamma) = \tr(Z|\x|^{-1} \gamma) -
D(\rho_\gamma,\rho_\gamma) + X(\gamma^{1/2})\ .
\end{equation}
\end{proposition}

\begin{proof}
For any $\lambda>0$ the density matrix $\gamma_\lambda$ defined by
$\gamma_\lambda(\X,\X') = \lambda^3
\gamma(\lambda\x,\sigma,\lambda\y,\sigma')$ is unitarily equivalent to
$\gamma$ and hence satisfies $0\leq\gamma_\lambda\leq 1$ and
$\tr\gamma_\lambda=\tr\gamma\leq N$. Since $\gamma$ is a minimizer, the
function
\begin{equation*}
\mathcal{\widehat
      E}^{\rm M}(\gamma_\lambda) = \lambda^2 \tr(-\half\nabla^2\gamma) -
\lambda \tr(Z|\x|^{-1} \gamma) + \frac 18 \tr \gamma + \lambda
D(\rho_\gamma,\rho_\gamma) - \lambda X(\gamma^{1/2})
\end{equation*}
has a minimum at $\lambda=1$. This implies the assertion.
\end{proof}

\section{The M\"uller Functional as a Lower Bound to Quantum Mechanics}

We are able to show that the M\"uller energy $E^{\rm M}(N)$
(without the addition of $N/8$) is a lower bound to the true Schr\"odinger energy
when $N=2$, but with arbitrarily many nuclei. The situation for $N>2 $ is
open. As we remark below, our $N=2$ proof definitely fails when $N>2$.

Consider the $N$-particle Hamiltonian (\ref{ham}) 
in either the symmetric or the anti-symmetric $N$-fold tensor product of
$L^2(\R^3,\C^q)$. For a symmetric or anti-symmetric $\psi$ we recall that
the one-particle density matrix $\gamma_\psi$ is defined by
$$
              \gamma_\psi(\X,\X') = N  \int
              \psi(\X,\X_2,\ldots,\X_N)\psi(\X',\X_2,\ldots,\X_N)^*
              \, d\X_2\cdots d\X_N \ .
$$

\begin{proposition}\label{lowerbound}
              Assume that $N=2$. Then for any symmetric or anti-symmetric
normalized $\psi$,
              \begin{equation*}
                      \bra\psi| H| \psi\ket \geq \mathcal E^{\rm
                        M}(\gamma_\psi)   \ .
              \end{equation*}
\end{proposition}

\begin{proof}
              Since $\bra\psi| \sum_{j=1}^2
(-\half\nabla^2_j- V_c(\x_j))|\psi\ket=
\tr(-\half\nabla^2-V_c(\x))\gamma_\psi$, we have
to prove that
              \begin{equation*}
                      \int \frac{|\psi(\X_1,\X_2)|^2}{|\x_1-\x_2|}\,d\X_1\,d\X_2+ \int
\frac{|\gamma_\psi^{1/2}(\X,\X')|^2}{2|\x-\y|}\,d\X\,d\X'
                      \geq \int \frac{\gamma_\psi(\X_1,\X_1) \gamma_\psi(\X_2,\X_2)}{2|\x_1-\x_2|}\,d\X_1\,d\X_2\ .
              \end{equation*}
              By \eqref{eq:fefferman} it suffices to prove that for any
              characteristic function $\chi$ of a ball (or, more
              generally, for any real-valued function $\chi$)
              \begin{equation}
                      2 \int \chi(\x_1) |\psi(\X_1,\X_2)|^2 \chi(\x_2)d\X_1d\X_2
+ \int \chi(\x)
|\gamma_\psi^{1/2}(\X,\X')|^2 \chi(\y)d\X d\X'\label{junk}
                      \geq \left(\int \chi(\x) \gamma_\psi(\X,\X) d\X\right)^2.
              \end{equation}
              Introducing $\Psi$ as the (non-self-adjoint) operator in $L^2(\R^3)$ with
kernel $\sqrt 2 \, \psi(\X,\X')$, we can rewrite the previous inequality as
              \begin{equation}\label{eq:lowerbound1}
                      \tr\chi\Psi^\dagger\chi\Psi + \tr\chi\gamma_\psi^{1/2}\chi\gamma_\psi^{1/2}
                      \geq (\tr \chi\gamma_\psi)^2 \ .
              \end{equation}
The proof of this inequality can be found in \cite{WignerYanase1963}.
For completeness, we present the proof here.

             Note that $\Psi\Psi^\dagger=\gamma_\psi$, so
$\Psi=\gamma_\psi^{1/2} \mathcal V$ for a partial isometry $\mathcal V$.
Since $\psi$ is (anti-) symmetric, $\Psi^\dagger\Psi=\mathcal
C\gamma_\psi\mathcal C$, where $\mathcal C$ denotes complex conjugation.
Hence $\mathcal V^\dagger \gamma_\psi \mathcal V = \mathcal
C\gamma_\psi\mathcal C$ and, since the square root is uniquely defined,
              \begin{equation}\label{eq:commutesqrt}
                      \mathcal V^\dagger\gamma_\psi^{1/2}\mathcal V = \mathcal C \gamma_\psi^{1/2}
\mathcal C \ .
              \end{equation}
              We write $\delta=\gamma_\psi^{1/2}$ for simplicity and consider the
quadratic form
              \begin{equation*}
                      Q(A,C)=
                      \frac14(2\tr A^\dagger\delta C\delta
                      + \tr A^\dagger\delta\mathcal V C\mathcal V^\dagger\delta
                      + \tr \mathcal V A^\dagger\mathcal V^\dagger\delta C \delta) \ .
              \end{equation*}
              We consider this quadratic form on the real vector space of \emph{real}
operators, i.e., operators satisfying
              \begin{equation}\label{eq:commuteop}
                      \mathcal C A \mathcal C =A \ .
              \end{equation}
              Note that $Q(A,A)= \frac12(\tr A^\dagger\delta A\delta + \tr A^\dagger\delta\mathcal
V A\mathcal V^\dagger\delta)$ and that, by Schwarz's inequality,
              \begin{equation*}
                      (\tr A^\dagger\delta\mathcal V A\mathcal V^\dagger\delta)^2
                      \leq (\tr A^\dagger\delta A\delta)
                      (\tr \mathcal V A^\dagger\mathcal V^\dagger\delta\mathcal V A\mathcal V^\dagger\delta) \ .
              \end{equation*}
              Recalling \eqref{eq:commutesqrt} and \eqref{eq:commuteop}
              we thus see that $Q$ is positive semi-definite. This
              implies in particular that $Q(\chi,1)^2\leq
              Q(\chi,\chi)Q(1,1)$. This is the desired inequality
              \eqref{eq:lowerbound1}, since $Q(1,1)=\tr\gamma_\psi=2$,
              $Q(\chi)=\frac12(\tr \chi\delta \chi\delta + \tr
              \chi\delta\mathcal V \chi\mathcal V^\dagger\delta) = \frac12(\tr
              \chi\delta \chi\delta + \tr \chi\Psi \chi\Psi^\dagger)$ and
              \begin{equation*}
                      Q(1,\chi)=
                      \frac14(3\tr \chi\delta^2
                      + \tr \chi \mathcal V^\dagger \delta^2\mathcal V ) = \tr \chi\gamma_\psi \ .
              \end{equation*}
              Here we used \eqref{eq:commutesqrt} once more.
\end{proof}

The obvious generalization of inequality \eqref{junk} to $N\geq 3$ is
not true, as the paper \cite{Seiringer2007} shows. But this does not
mean that the M\"uller energy is not a lower bound to the true energy.
There is some
numerical evidence for this, as mentioned
in A.3 of subsection \ref{1B}.

\bigskip

\underline{\textit{Acknowledgments:}} Rupert Frank and Heinz Siedentop thank the
Departments of Mathematics and Physics at Princeton University for
hospitality while this work was done.  The following partial support is 
gratefully acknowledged:  The Swedish Foundation for International Cooperation
in Research and Higher Education (STINT) (R.F.);
U.S. National Science Foundation, grants PHY 01 39984 (E.H.L and H.S.) 
and  PHY 03 53181 (R.S.); an A.P. Sloan Fellowship (R.S.);
Deutsche Forschungsgemeinschaft, grant SI 348/13-1 (H.S.).



\end{document}